\documentclass[reprint,superscriptaddress,amsmath,amssymb,prc,USenglish,floatfix]{revtex4-2}
\usepackage{babel}
\usepackage{graphicx}
\usepackage{booktabs}
\usepackage[separate-uncertainty=true]{siunitx}
\usepackage{tensor}
\usepackage{braket}
\usepackage{tikz}
\usepackage{cleveref}
\renewcommand{\d}{\mathop{}\!\mathrm{d}}

\newcommand{\ccite}{ref.~\cite}
\newcommand{\ccites}{refs.~\cite}
\newcommand{\nnom}{l}  \newcommand{\ctom}{\ell}    \newcommand{\PTom}{L}  \newcommand{\totAM}{J}  \newcommand{\totP}{\pi}  \newcommand{\npPair}{\nuclide{n}\nuclide{p}} \newcommand{\npPairGroup}{(\nuclide{n}\nuclide{p})} \newcommand{\npSystem}{\nuclide{n}--\nuclide{p}}

\begin{document}
\title{Clustering effects in the \nuclide[6]{Li}(\nuclide{p},\nuclide[3]{He})\nuclide[4]{He} reaction at astrophysical energies}
\author{S.~S.~Perrotta}%
\email{perrotta@lns.infn.it}\altaffiliation{\\Current address: Lawrence Livermore National Laboratory, 94551 Livermore, CA, United States.}%
\affiliation{Dipartimento di Fisica e Astronomia, Università degli Studi di Catania,  I-95123 Catania, Italy}%
\affiliation{Departamento de Física Atómica, Molecular y Nuclear, Universidad de Sevilla, Apartado 1065, E-41080 Sevilla, Spain}%
\affiliation{Laboratori Nazionali del Sud, Istituto Nazionale di Fisica Nucleare, I-95123 Catania, Italy}
\author{M.~Colonna}
\email{colonna@lns.infn.it}
\affiliation{Laboratori Nazionali del Sud, Istituto Nazionale di Fisica Nucleare, I-95123 Catania, Italy}
\author{J.~A.~Lay}
\email{lay@us.es}
\affiliation{Departamento de Física Atómica, Molecular y Nuclear, Universidad de Sevilla, Apartado 1065, E-41080 Sevilla, Spain}
\affiliation{Instituto Interuniversitario Carlos I de Física Teórica y Computacional (iC1), Apdo. 1065, E-41080 Sevilla, Spain}
\date{\today}
\begin{abstract}\begin{description}
    \item[Background] The understanding of
    nuclear reactions between light nuclei at energies below the Coulomb barrier is important for several astrophysical processes, but their study
    poses experimental and theoretical challenges. At sufficiently low energies, the electrons surrounding the interacting ions affect the scattering process. Moreover,
    the clustered structure of some of these nuclei may play a relevant role on the reaction observables.
    \item[Purpose] In this article, we focus on a theoretical investigation of
    the role of clustered configurations of \nuclide[6]{Li} in reactions of astrophysical interest.
    \item[Methods] The \nuclide[6]{Li}(\nuclide{p},\nuclide[3]{He})\nuclide[4]{He} reaction cross section is described considering both the direct transfer of a deuteron as a single point-like particle in Distorted Wave Born Approximation (DWBA), and the transfer of a neutron and a proton in second-order DWBA. A number of two- and three-cluster structure models for \nuclide[6]{Li} are compared.
    \item[Results] Within the two-cluster structure model, we explore
    the impact of the deformed components in the \nuclide[6]{Li} wave-function on
    the reaction of interest. Within the three-cluster structure model, we gauge the degree of \nuclide{\alpha}--\nuclide{d} clustering and explicitly probe its role on specific features of the reaction cross section. We compare the energy trend of the astrophysical $S$ factor deduced in each case.
    \item[Conclusions] Clustered \nuclide[6]{Li} configurations lead in general to a significant enhancement of the astrophysical factor in the energy region under study. This effect only originates from clustering, whereas static deformations of the ground-state configuration play a negligible role at very low energies.
\end{description}\end{abstract}
\maketitle
\section{Introduction}
 The characterization of reaction mechanisms occurring at low energies, 
and in particular below the Coulomb barrier, represents one of the most interesting challenges of the last decades for nuclear physics.
More specifically, reactions involving light nuclei are experiencing renewed
interest, also for their important implications in the astrophysical context \cite{rolfs1988cauldrons,ThompsonNunes2009,Adelberger2011,Bertulani2016,Lamia2013,Spitaleri2016,Fang2018,Typel2020}.

While nuclear reactions at energies in the range $\geq$~1~MeV are crucial for
explosive-type evolutionary processes in
the Universe, the lower energy domain  ($\le$ 200-300 keV) is relevant
to the study of the nucleosynthesis processes taking place shortly after the Big Bang and in quiescent-phase stars.   

Low energy nuclear reactions essentially include transfer and capture processes. The former are mainly governed by the 
nuclear force, whereas the latter are mainly driven by the electromagnetic 
interaction. The determination of cross sections at stellar energies, 
which are in general much lower than the Coulomb barrier, 
requires considerable efforts.
 Indeed, in most cases, direct measurements
are really difficult because, for energies within the Gamow window, 
the corresponding cross sections
become extremely low (reaching the nano or picobarn regime) \cite{rolfs1988cauldrons,Broggini2010}.
 Although new experimental techniques, including indirect 
measurement methods, have been  developed over the last decades \cite{Tribble14},  the extrapolation of data down to stellar energies often requires theoretical insight.

Quantum scattering theory provides solid foundations for cross-section calculations.
A variety of models, well suited to the low energies (and low level density of the {involved} light nuclei)
relevant for nuclear astrophysics, are widely employed in current studies, such as, for instance, optical models for capture reactions~\cite{Bertulani2003}  and (first- and second-order) DWBA and/or coupled-reaction-channel calculations for transfer
reaction mechanisms.  Within such a context, the phenomenological R-matrix theory (see for example \ccites{barker199112c,Descouvemont2010})  has proven to be an efficient tool to investigate reactions of astrophysical interest, using  existing data at higher energies as a reference point.
Major recent developments also concern the formulation of more microscopic 
(ab-initio) models, such as the Resonating Group Method~\cite{Arai2002}, relying only on the bare nucleon-nucleon (n-n) interaction, with few or no adjustable parameters, thus having in principle high predictive power. These studies are of wide interest, also for the possibility to probe yet unknown aspects of the 
n-n interaction and explore derivation schemes involving sub-nucleonic degrees of freedom
\cite{Gnech2019,Gnech2020}.
However, solving the many-body Schrödinger equation for scattering states is a quite difficult task that  can only be accomplished  for specific reactions
involving relatively light nuclei. Suitable approximate methods, such as 
the cluster approximation~\cite{horiuchi2012recent}, are often employed.

The possible influence of clustering effects,  
characterizing the structure of light nuclei, 
on reaction mechanisms occurring at very low energies deserves particular 
attention. Several light nuclei of relevant interest in the astrophysical context, like 
 \nuclide[6,7]{Li}, \nuclide[7,9]{Be} 
and \nuclide[10,11]{B}, are known to exhibit a pronounced cluster structure.
Particularly important is the case of \nuclide[6]{Li}, because its abundance in the stellar environment constrains the \nuclide[7]{Li}-depleting mechanisms and their efficiency. 
Current stellar models are unable to predict the observed
surface lithium abundance. This is often referred to as the “lithium problem", which has stimulated several investigations,
both on the theoretical and experimental side, aiming at improving the accuracy of low-energy bare-nucleus cross sections of lithium-burning reactions (see \cite{Lamia2013} and refs. therein). Reactions destroying \nuclide[6]{Li} are also interesting for studying pre-main-sequence  stars. 

It is well known that low-energy fixed-target
direct measurements exhibit a cross
section enhancement because of the electron screening effect caused by the electron clouds surrounding the interacting ions. 
This phenomenon makes it more difficult to measure
the bare-nucleus cross section \cite{Assenbaum1987,strieder2001electron},  which is needed to subsequently evaluate the cross section relevant for astrophysical environments.  An anomalous cross-section enhancement, larger than predicted by standard electron-screening calculations,  has been experimentally detected in several reactions 
at astrophysical energies (an overview can be found in \ccites{Fiorentini2003,Spitaleri2016}).
This observation, 
known as the electron screening problem, calls for a deeper analysis of nuclear reaction rates at stellar energies, with particular reference to the possible impact of clustering effects~\cite{Spitaleri2016}.

In this work, we undertake a theoretical investigation of the 
$\nuclide[6]{Li} + \nuclide{p} \rightarrow \nuclide{\alpha} + \nuclide[3]{He}$ reaction cross-section at energies below and around the Coulomb barrier, within the framework of first and second-order DWBA. We focus on the impact of the reactants' structure,
and in particular of the \nuclide[6]{Li} ground state,  by considering different configurations and investigating their influence on the transfer cross section, for a fixed set of optical and binding potentials.  The main goal of this analysis is to investigate a possible sensitivity of the results  to clustering and static deformation effects. Thus, explicit “dynamical” effects, namely the coupling to excited states or other competing channels, are not included here and
will be the subject of future investigations.  

The paper is organized as it follows: in section~\ref{secPhenomenologicalAnalysis}  we discuss in greater detail  the electron screening problem and its impact  on the astrophysical $S$-factor.  The methodology adopted for the transfer calculations, together  
with the details referring to optical and binding potentials and to
structure inputs, is described in section~\ref{secMethodology}.  The results obtained for the transfer cross section, and associated astrophysical factor, as a function of the beam energy are presented in section~\ref{secResults}.  Conclusive remarks and perspectives are given in section~\ref{secConclusions}.  
\section{Phenomenological analysis of reactions at stellar energies}\label{secPhenomenologicalAnalysis}

In this section, we will review some  techniques and phenomenological methods  employed to analyze reaction cross sections at stellar energies, with particular reference to the $\nuclide[6]{Li} + \nuclide{p} \rightarrow \nuclide{\alpha} + \nuclide[3]{He}$ reaction, for which several sets of experimental data are available in literature. 

 \subsection{Low-energy reactions and screening effects}

Let us consider a reaction between two nuclei with charge numbers $Z_1$ and $Z_2$ and reduced mass $m$.  Denoting by $\sigma(E)$ the non-polarized angle-integrated cross section at a given center-of-mass collision energy $E$, the corresponding astrophysical factor, $S(E)$,
is defined as (see e.g.~\cite[eq.~1.1.4]{ThompsonNunes2009})
	 \begin{equation}\label{eqDefAstrophysicalFactor}
S(E) = E \, e^{2 \pi \eta(E)} \, \sigma(E) , \quad
\eta(E) = \alpha_e Z_1 Z_2 \sqrt{\frac{m}{2 E}},
\end{equation}
	 where $\alpha_e\approx 1/137$ is the fine-structure constant and $\eta$ denotes the Sommerfeld parameter.  
 The astrophysical $S$-factor is a  convenient tool to represent and discuss results at sub-Coulomb  energies, as a relevant portion of the exponential dependence on energy of the cross-section is factored out.  Thus, in the following we will adopt it for our analysis of transfer cross sections.  
 
 At very low energies, screening effects associated with the electrons surrounding the interacting ions  become rather important. Since the  effects found in laboratory experiments are of different kind and amplitude as compared to those appearing in astrophysical environments,  information on the scattering between isolated reactants   is usually required for the modeling of astrophysical reactions. 
An interesting possibility to get around this problem is represented by data
obtained with indirect measurements, like the ones based on the
Trojan Horse Method (THM) \cite{Lamia2013}, which are not affected by electron screening and thus  can potentially provide information on the bare-nucleus cross section even at vanishing beam energy.
At the same time, 
for fixed-target direct measurements such as the one  reported in \ccite{Elwyn1979}, one can safely  
 rely on data  at sufficiently high collision energies,  where electrons play a negligible role.

Many efforts have been devoted in the past to the possibility to extract the bare cross
section, also at very low energies, from direct measurements (see \ccites{Assenbaum1987,Barker2002,Cruz2005} for some examples), thus identifying  the screening effects. 
  	Here, we will consider low-energy direct data      in \ccites{Engstler1992,CruzPhD2006}
	 	 	measured  	employing targets in which electrons can be approximated as belonging   only     to atomic (or molecular) bound states.  The effects  in the low-energy reaction cross-section,   caused by interactions between the reactants and the surrounding environment, 
are  commonly described using the screening potential approximation. Within the context of fixed-target nuclear reaction experiments, the approach is described in \ccite{Assenbaum1987} and treated in-depth in \ccite{Bracci1990};  the same model was employed since much earlier to describe screening effects in plasma environments \cite{Salpeter1954}. The approximation prescribes to express the screened cross section, $\sigma_s$, in terms of the bare-nucleus one, $\sigma_b$, as
\begin{equation}\label{eqDefScreeningPotentialApproximation}
    \sigma_s(E) = \sigma_b(E + U),
\end{equation}
 where $U$ denotes the screening potential, which is often approximated to a constant with respect to energy.   In \ccite[eq.~(2.17)]{Bracci1990} it is estimated that using such formalism will induce a relative error of the order of $U/E$ on screened cross-sections of atomic systems.
  In \ccite[sec.~5]{Bracci1990}, atomic screening effects are discussed in the limit of small collision energies, under the so-called adiabatic limit. If the initial and final state for the atomic systems is the ground state (with no degeneracy), the adiabatic limit yields  the theoretical upper limit for the screening potential at zero collision energy, which is just the difference between the final (corresponding to the
touching-point configuration) and initial binding energy for the projectile and target electron systems (namely, the maximum amount of energy that electrons can release to the nuclear-motion degrees of freedom).  A system made of a neutral hydrogen atom impinging on a neutral \nuclide{Li} atom has an adiabatic-limit screening potential of \SI{182}{\eV} \cite[tab.~4]{Bracci1990}.
 
Substituting \cref{eqDefAstrophysicalFactor} into \cref{eqDefScreeningPotentialApproximation}, the corresponding screened astrophysical factor, $S_s$, is consequently connected to  the bare one, $S_b$, as follows:
 \begin{equation}\label{eqEnhancementFactorConstantSfactor}
         S_s(E) = \frac{E}{E+U} \, e^{2 \pi \left[ \eta(E) - \eta(E+U) \right]} S_b(E+U)           .
	\end{equation}
 In this expression,  it is often possible to approximate $S_b(E+U) \approx S_b(E)$, since the astrophysical factor likely varies very slowly within an energy range equal to the typical values of $U$ (few hundreds of \unit{\eV} at most for the systems here of interest). \Cref{eqEnhancementFactorConstantSfactor} can then be employed to gauge the ratio $f_e = \sigma_s(E) / \sigma_b(E)$, namely the \emph{enhancement factor}.
                     For instance,
	assuming $U = \SI{182}{\eV}$,
	 	 	the quantity $[f_e(E)-1]$ for a $\nuclide[6]{Li} + \nuclide{p}$ reaction  	at center-of-mass energies above \SI{75}{\keV}
	is approximately \SI{1}{\percent} or smaller, well below typical experimental errors.  On the other hand, the correction becomes significant at vanishing energies, 
for instance it is $[f_e(E)-1] \approx \SI{26}{\percent}$ at $E = \SI{10}{\keV}$ and $U$ as above.

\begin{figure}[tbp]
    \centering
\includegraphics[keepaspectratio=true, width=\linewidth]{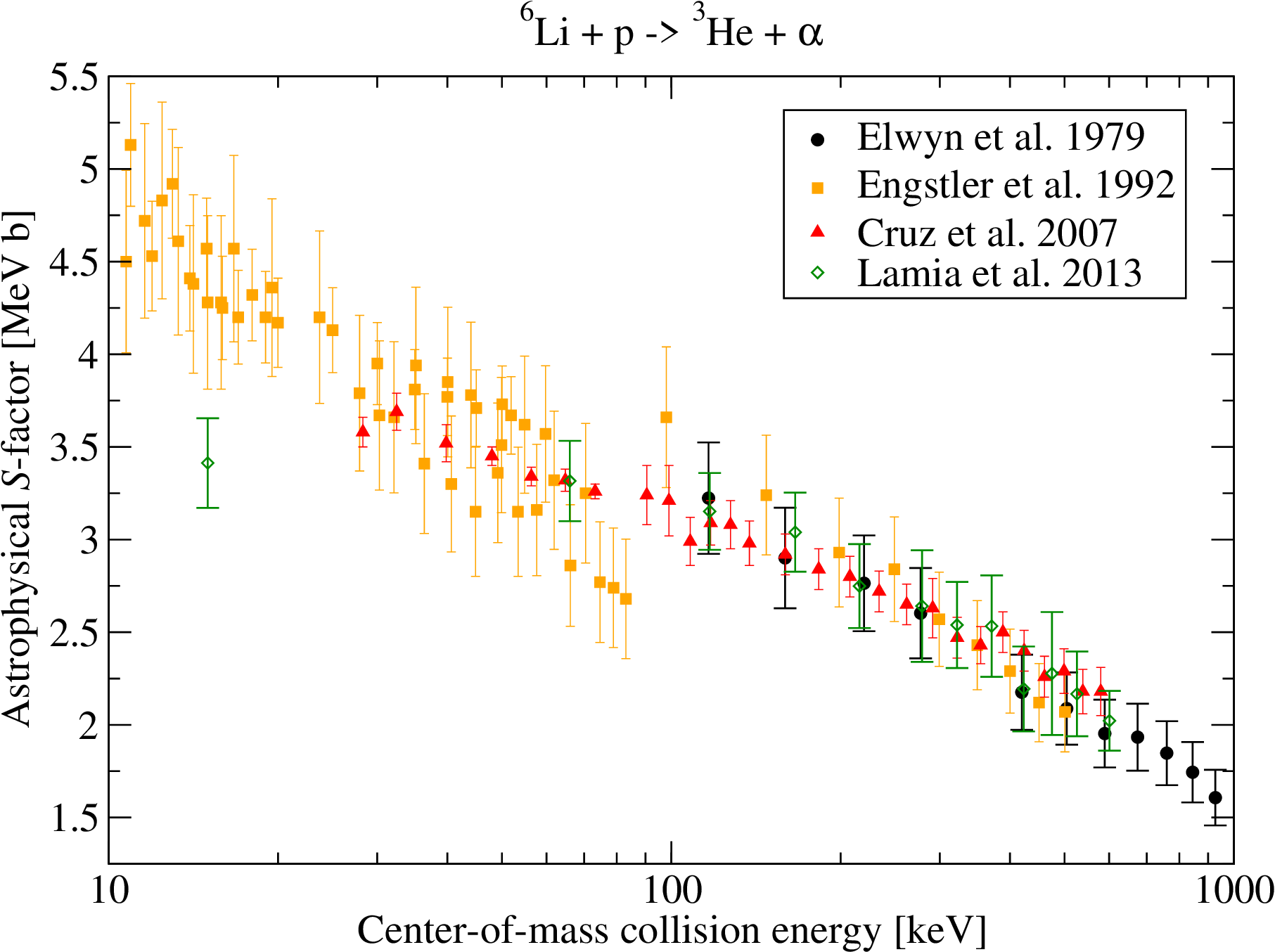}
    \vskip 1ex
\includegraphics[keepaspectratio=true, width=\linewidth]{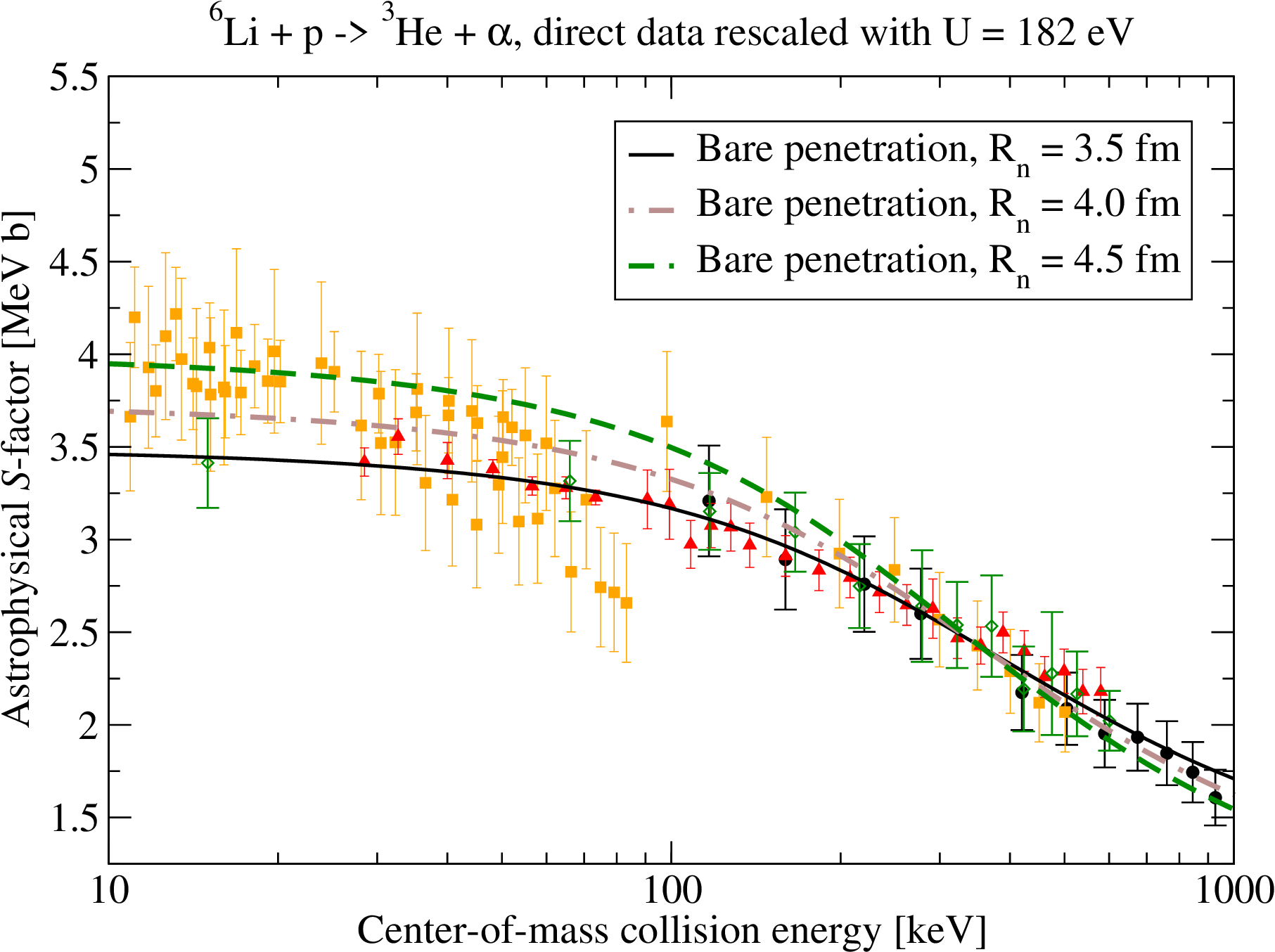}
    \caption{Top panel:      Experimental $\nuclide[6]{Li}+\nuclide{p}\to\nuclide[3]{He}+\nuclide{\alpha}$ astrophysical factors as a function of the collision energy, in particular      data below \SI{1}{\MeV}      from \ccite{Elwyn1979} (black circles), inverse-kinematics data 
    from \ccite{Engstler1992} (orange full squares), data from \ccite{Cruz2007} including insulator-target data      first published in \ccite{Cruz2005} (red upward triangles),      and Trojan Horse Method data from \ccite{Lamia2013} (green open diamonds).\\Bottom panel: green open diamonds are the same as in the top panel. All other points represent the predicted bare-nucleus astrophysical factor obtained from data in the top panel using \cref{eqDefScreeningPotentialApproximation} with $U = \SI{182}{\eV}$. Each line 
    corresponds to \cref{eqCrossSectionPenetrability} with a given value of $R_n$, as per the legend, and $C$ adjusted so that the line agrees with experimental data at \SI{350}{\keV} (see text).}
    \label{fig1}
\end{figure}

\Cref{fig1} illustrates the effect of the expected electron screening in the case of the $\nuclide[6]{Li} + \nuclide{p} \rightarrow \nuclide{\alpha} + \nuclide[3]{He}$ reaction. The top panel of \cref{fig1} compares the bare-nucleus astrophysical factor from ref.~\cite{Lamia2013}, measured using the Trojan Morse Method, with screened astrophysical factors obtained through direct measurements reported in \ccites{Elwyn1979,Engstler1992,Cruz2005,Cruz2007}, showing a clear enhancement toward low energies.
The bottom panel of \cref{fig1} instead reports the prediction for the corresponding bare-nucleus astrophysical factor, obtained using   \cref{eqDefScreeningPotentialApproximation} 
with  $U = \SI{182}{\eV}$. 
The impact of screening effects on the measured astrophysical factor is nicely evidenced by this simple approach.
  As can be seen from the figure, a screening potential of \SI{182}{\eV} largely removes the difference in energy 
trend between direct and indirect data discussed above. 
However, in spite of the large error bars, the direct data (see in particular the points from ref.~\cite{Engstler1992}) still exhibit a clear enhancement at very low energies.
This effect, namely the “electron screening problem", could be attributed
to nuclear features (such as clustering or deformation effects) which could further reduce the Coulomb repulsion between the reactants, as we will investigate in the following. 
  
\subsection{Phenomenological description of the cross section}\label{secIntroPhenomenologicalPenetrabilityModel}
    
     	At sufficiently low energies,  	the $\nuclide[6]{Li} + \nuclide{p} \rightarrow \nuclide{\alpha} + \nuclide[3]{He}$ bare-nucleus  	astrophysical factor decreases approximately linearly with energy, hence in the bottom panel of \cref{fig1} one observes a rather flat behavior as a function of $\ln(E)$,  	suggesting the absence of measurable contributions from resonances.
	It is interesting to investigate up to which extent this trend resembles the one expected from a phenomenological Coulomb-barrier-penetrability model. 
  
  At energies $E$ well below the reactants Coulomb barrier, the reaction dynamics tends to be dominated by the process of quantum tunneling through the barrier.
  	Considering for simplicity a sharp-edge spherical well
	for the projectile-target nuclear
	interaction,
	the angle-integrated reaction cross-section $\sigma_b(E)$ for the process of Coulomb-barrier penetration can be expressed as in ref.~\cite[sec.~1.1.1]{PerrottaTesiPhD}. As a matter of fact, only channels with small projectile-target orbital angular momentum are relevant, because the centrifugal barrier further hinders penetrability.
	Hence, considering only $s$-wave scattering, one can write:  	\begin{equation}\label{eqCrossSectionPenetrability}
	\sigma_b(E) \approx C \frac{\pi}{k^2} P_0,
	\qquad
	P_0 = \frac{k R_n}{|H_0(\eta, k R_n)|^2},
	\end{equation}
	 	where $k = \sqrt{2 m E} / \hbar$ is the entrance channel relative momentum,
	and $P_0$ is the penetrability factor found also in R-matrix theory~\cite[eq.~(10.2.5)]{ThompsonNunes2009}.
    $C$ is a dimensionless free parameter, phenomenologically accounting for reaction details such as the overlap between initial and final states and phase-space factors.
 $H_l(\eta, k r)$ is the spherical Coulomb wave-function {\cite[ch.~33]{NISTdlmf}} and $R_n$ is a parameter representing the distance between the two nuclei at the contact point, usually taken equal to 
 $R_0 = \qty{1.2}{\femto\metre}~(A_1^{1/3} + A_2^{1/3})$, which is about \qty{3.5}{\femto\metre} for the reaction considered here.
    In the sharp-edge spherical well model, $R_n$ also represents the radius at which the potential reaches its maximum (the barrier height),
    $k_e Z_1 Z_2 / R_n$ (where $k_e \approx \qty{1.44}{\MeV \femto\metre}$), which in the present case is about \qty{1.3}{\MeV} for $R_n = R_0$.
    
    The curves reported in \cref{fig1} (bottom panel)  have been obtained considering several values for $R_n$ and adjusting $C$ to fit experimental data at about \SI{350}{\keV}.  
One can observe that, whereas the THM indirect data are nicely fitted by $R_n = R_0$, a larger $R_n$ value is needed to reproduce the scaled direct measurements at the lowest energies.
 This can be taken as an indication of the fact that the experimental cross section would be compatible with a lowering of the Coulomb barrier, possibly induced
by the presence of clustering or deformation effects in the \nuclide[6]{Li} nucleus \cite{Spitaleri2016}. 
     In the following, we will explore the possible role of such effects in the $(\nuclide{p},\nuclide[3]{He})$  transfer process.

\section{Methodology}\label{secMethodology}
When studying (\nuclide{p},$^3$He) reactions, one can take advantage of the fact that 
the transferred system, namely a deuteron (\nuclide{d}), is bound and assume that the process corresponds to the transfer of an elementary particle. In such a case, the cross section can be obtained within the first-order Distorted-Wave Born Approximation (DWBA) \cite[sec.~2.8]{Satchler1983Direct}. One can also go beyond this approximation by assuming that the reaction mechanism involves the transfer of two individual nucleons, 
namely one neutron and one proton (\npPair{} transfer). Then the reaction process acquires a greater degree of complexity and needs to be treated in second-order DWBA~\cite{Satchler1983Direct,Thompson2013}.    Recently, there has been a revival in the use of second-order DWBA  thanks to its success in the study of two-neutron pairing correlations, providing absolute cross sections in very good agreement with the data~\cite{Potel2013,Potel2013b,Tanihata2008,Des21}. 
The \npPair{} transfer is a less explored case. To our knowledge, the first attempt to perform second-order DWBA calculations for a \npPair{} transfer reaction was discussed
in ref.~\cite{Ayyad2017}, to investigate the relative importance of isoscalar (total isospin $T=0$ for the \npPair{} pair) and isovector ($T=1$) pairing.

    In the following, the basic formalism and the different ingredients of the calculations performed in the present study are introduced. Consider a generic reaction $\mathcal{A}+b \to a + \mathcal{B}$ proceeding as the transfer of the system $N$,      meaning that      $\mathcal A$ (or $\mathcal B$) can be seen as a bound state of particles      $a$ (or $b$) and $N$.      Let $\phi_i$ be the wave-function describing the internal motion of      particle $i$ in the state of interest for the process.      For the reaction considered in the present work, the initial system involves \nuclide{p} and \nuclide[6]{Li} in its ground state, and the final one comprises the ground states of \nuclide[4]{He} and \nuclide[3]{He}.

                             The complete model Hamiltonian $\mathcal H$ can then be decomposed as           \begin{equation}\label{eqHamiltonianOneParticleTransfer}
        \mathcal H = H_0 + \sum_{i\in\lbrace a,b,N\rbrace} K_i + V_{ab} + V_{aN} + V_{bN} ,
             \end{equation}
         where $H_0$ describes the internal motion of $a$, $b$ and $N$,      $K_i$ is the kinetic energy of the center of mass of      particle $i$, and $V_{ij}$ is the potential between particles $i$ and $j$.                       The internal-state wave-function for the initial partition, $\phi_{\mathcal A} \phi_b$, will be an eigenfunction of $H_0 + K_{aN} + V_{aN}$, where $K_{ij}$ is the kinetic energy for the relative motion between the centers of mass of $i$ and $j$. Analogous considerations apply to $\phi_a \phi_{\mathcal B}$.

\subsection{One-particle transfer}\label{secMethodologyOPT}

         The one-step transition amplitude for the reaction, $\mathcal T$, (whose square-modulus is proportional to the differential cross-section \cite[eq.~(12)]{MoroVarenna}) can be expressed in DWBA as in \ccite[eqs.~(2.96), (2.98)]{Satchler1983Direct}, in the so-called \emph{prior} or \emph{post} forms:        \begin{equation}\label{eqTAmplitudeOneparticleTransfer}\begin{aligned}
    {\mathcal T}^{\text{prior}} &= \Braket{ \chi_{a\mathcal B}^{(-)}\phi_{a}\phi_{\mathcal B} | V_{bN} + V_{ab} - U_{\mathcal A b} | \phi_{\mathcal A} \phi_{b} \chi_{\mathcal A b}^{(+)} } , \\
    {\mathcal T}^{\text{post}} &= \Braket{ \chi_{a\mathcal B}^{(-)}\phi_{a}\phi_{\mathcal B} | V_{aN} + V_{ab}-U_{a\mathcal B} | \phi_{\mathcal A} \phi_{b} \chi_{\mathcal A b}^{(+)} } ,
\end{aligned}\end{equation}
         where $U_{\mathcal A b}$ and $U_{a \mathcal B}$ are initial- and final-state optical potentials, acting only on the relative motion between the centers of mass of projectile and target in the respective partitions;
    $\chi_{\mathcal A b}^{(+)}$ and $\chi_{a\mathcal B}^{(-)}$ are distorted waves, namely eigenfunctions of $U_{\mathcal A b}$ and $U_{a \mathcal B}$      with boundary conditions appropriate for the problem of interest.
              If no further approximations are introduced, the results do not depend on the chosen form (prior or post) \cite[sec.~2.8.8]{Satchler1983Direct}.

    In the one-particle-transfer description, the reaction takes place as the direct transfer of an inert system, $N$, which is the ground state of a free deuteron in the present calculation.      The complete internal state of $\mathcal A$ is truncated to a single component factorized in the internal state of the free particles $a$ and $N$, and a bound wave-function for the motion of the centers of mass of $a$ and $N$, $\phi_{aN}$, obtained from the overlap function $\Braket{\phi_{a} \phi_N|\phi_{\mathcal A}}$. The same applies to the internal state of $\mathcal B$. The particles $a$, $b$ and $N$ are thus effectively treated as inert and point-like, apart from the appearance of the spectroscopic amplitudes associated with the overlap functions.
         The factorized form of the wave-functions allows to simplify the matrix elements in \cref{eqTAmplitudeOneparticleTransfer} performing the integrals over all internal-motion coordinates of $a$, $b$, $N$. For instance, the term $\Braket{\phi_{a}\phi_{\mathcal B} | V_{bN} + V_{ab} | \phi_{\mathcal A} \phi_{b}}$ can be written as $\Braket{ \phi_{bN} | \mathcal V_{bN} + \mathcal V_{ab} | \phi_{aN} }$, where the potentials $\mathcal V_{ij}$ describe the interaction between the centers of mass of particles $i$ and $j$ and do not act on their internal degrees of freedom.
    In practice, the many-body interactions $V$ are not introduced explicitly: the potentials $\mathcal V_{aN}$ and $\mathcal V_{bN}$ are set so that $\phi_{aN}$ and $\phi_{bN}$ are eigenfunctions of such potentials with the desired separation energy as eigenvalue, and the core-core interaction $\mathcal V_{ab}$ is defined phenomenologically.

\subsection{Two-particle transfer}\label{secMethodTwoParticleTransfer}
    In the two-particle-transfer formulation, the transferred system, $N$, is explicitly modeled as composed by two sub-systems, $\mu$ and $\nu$, here a proton and a neutron. $\mathcal{A}$ is thus the composition of $a$, $\mu$ and $\nu$, and similarly $\mathcal{B} = b+\mu+\nu$.
    Additionally, the model space is extended to include an intermediate state where only particle $\nu$ is transferred, comprising the nuclei $A = a + \mu$ and $B = b + \nu$.      The total Hamiltonian of the system in \cref{eqHamiltonianOneParticleTransfer} can now be written more explicitly:  	     \begin{equation}\label{eqHamiltonianTNT}
    \mathcal H = \tilde H_0 + \sum_{i\in\lbrace a,b,\mu,\nu\rbrace} K_{i} + V_{ab} + V_{a\mu} + V_{a\nu} + V_{b\mu} + V_{b\nu} + V_{\mu\nu} ,
    \end{equation}
         where $\tilde H_0$ describes the internal motion of $a$, $b$, $\nu$ and $\mu$.
    As in the one-particle-transfer case (see again sec.~\ref{secMethodologyOPT}), the many-body interactions $V_{ij}$ will not be explicitly employed in the practical calculations, since all matrix elements of interest can be simplified integrating over the internal coordinates of $a$, $b$, $\nu$ and $\mu$.
    
    The process was treated in second-order DWBA. 
              The reaction can proceed as a one-step transfer from $\mathcal{A}+b$ to $a+\mathcal B$ as in sec.~\ref{secMethodologyOPT}, which corresponds to a transition matrix, $\mathcal T^{(1)}$, denoted as \emph{simultaneous} term: this is formally identical to the one in \cref{eqTAmplitudeOneparticleTransfer} but keeping in mind that $V_{aN} = V_{a\nu} + V_{a\mu}$ (and similarly for $V_{bN}$), and that the system has now an extra degree of freedom (the internal motion within $N$).
    The final state of interest can also be reached      through a two-step process involving two single-particle      transfers of a nucleon (each, once again, analogous to the one described in sec.~\ref{secMethodologyOPT}) and the propagation of the intermediate state, $A+B$, populated by the first step of the process.
    In literature, the corresponding transition matrix (which is the same appearing in \ccite[eq.~(3.62)]{Satchler1983Direct}), $\mathcal T^{(2)}$, is often split into two terms
         denoted as \emph{sequential} and \emph{non-orthogonality} contributions. Here, the two latter terms are always considered together.           The total transition amplitude for the process is thus expressed as $\mathcal T = \mathcal T^{(1)} + \mathcal T^{(2)}$.
         Within DWBA, if no further approximations are introduced and provided that non-orthogonality terms are correctly taken into account, each transfer step appearing in both $\mathcal T^{(1)}$ and $\mathcal T^{(2)}$ can be indifferently computed in either the prior or post form \cite[sec.~3]{Thompson2013}.

\subsubsection{Simultaneous two-nucleon transfer}\label{secMethodTNTSimultaneous}
    The wave-functions of $\mathcal A$ and $\mathcal B$,      here labeled $\Psi_{\mathcal A}$ and $\Psi_{\mathcal B}$, are constructed in terms of a superposition of states factorized in the core-nucleon motions (“V coordinates"), for instance:   \begin{equation}\label{eqSuperpositionXCoordinates}
    \Psi_{\mathcal A}(r_{a\mu},r_{a\nu}) = \sum_{i,j} c_{i,j} \, \phi_{a\mu,i}(r_{a\mu}) \phi_{a\nu,j}(r_{a\nu}) \ ,
\end{equation}
which, if the isospin formalism is adopted,  needs to be properly antisymmetrized.     The state is such that the separation energy of $\mathcal A$ into $a$, $\mu$, and $\nu$ is fixed to the experimental value (this is necessary to ensure that $\mathcal A$ as a whole has the correct binding energy, or mass), and similarly for $\mathcal B$. This amounts to a constraint on the binding potentials of the system, which in practice is enforced by uniformly rescaling the potentials' volume term.

In principle, the functions $\phi$ in \cref{eqSuperpositionXCoordinates} are to be chosen so that $\Psi_{\mathcal A}$ is an actual eigenstate of the Hamiltonian for the isolated system $\mathcal A$, which is 
 \begin{equation}
    \sum_{i \in a,\mu,\nu} K_i + V_{a\mu} + V_{a\nu} + V_{\mu\nu} .
\end{equation}
      	     The total Hamiltonian in \cref{eqHamiltonianTNT}      can be recast in several possible ways      by regrouping the kinetic energy operators and
    moving to the center-of-mass frame of the whole $a+b+\nu+\mu$ system.
     	For instance,
	       \begin{subequations}\label{eqTwoNucTranHamiltExplicit}
\begin{multline}
	\mathcal H = \tilde H_0 + \left(K_{a\mu} + V_{a\mu}\right) + \left(K_{A\nu} + V_{a\nu} + V_{\mu\nu}\right) +\\+ \left(K_{\mathcal{A}b} + V_{ab} + V_{b\mu} + V_{b\nu} \right)      ~~~\text{(``prior''),}
\end{multline}
  \begin{multline}
	\mathcal H = \tilde H_0 + \left( K_{b\nu} + V_{b\nu} \right) + \left( K_{B\mu} + V_{b\mu} + V_{\mu\nu} \right) +\\+ \left(K_{a\mathcal{B}} + V_{ab} + V_{a\mu} + V_{a\nu} \right)      ~~~\text{(``post'').}
\end{multline}
\end{subequations}
	  where $K_{ij}$ is the kinetic energy for the relative motion between the centers of mass of $i$ and $j$.    	 	 	  $K_{a\mu} + V_{a\mu}$ can be seen as the ``internal'' Hamiltonian of the isolated $a+\mu$  system in its center-of-mass frame. Similarly, $K_{A\nu} + V_{a\nu} + V_{\mu\nu}$ is the Hamiltonian of the $A+\nu$ system, in its center-of-mass, deprived of the contribution from the internal motion of $A$.
 	       All the other terms enclosed in braces in \cref{eqTwoNucTranHamiltExplicit} bear an analogous meaning.
	Note      that $V_{ab} + V_{b\mu} + V_{b\nu}$ (the potential between $b$ and $\mathcal A$) is the potential appearing in the prior-form transition operator for the simultaneous-transfer calculation (the analogous of \cref{eqTAmplitudeOneparticleTransfer}).
    
	When practically performing two-nucleon-transfer simultaneous calculations, for reactions involving heavy ions, the Hamiltonian is often (see e.g.~\ccites[sec.~7]{Thompson2013}, \cite{Tanihata2008} and \cite{FrescoExample11Li})      approximated by formally setting, for either or both the projectile and target systems, $V_{\mu\nu} = 0$  	and using $K_{a\nu}$ and/or $K_{b \mu}$ in place of, respectively, $K_{A\nu}$ and $K_{B\mu}$. 
         This is generally accurate if      the core nucleus ($a$ or $b$) is much heavier than the transferred particles ($\mu$ and $\nu$).           Under such      approximation, here labeled as ``heavy-ion scheme'' for definiteness,                               the $\phi_{a\mu}$ and $\phi_{a\nu}$ in \cref{eqSuperpositionXCoordinates} become just the eigenstates of the $a$--$\mu$ and $a$--$\nu$ systems.
     	Additionally, the approximated Hamiltonian is such that the potential appearing in the prior-form simultaneous transition operator coincides with the sum of the core-core interaction, $V_{ab}$, and of the total binding potential of $\mathcal B$ (and similarly in post): this is the same condition found  	in single-particle transfers, and simplifies the practical calculation of the transfer cross-section.  	 	However, the kinetic terms in the approximated Hamiltonians in prior and post form are different, breaking the formal equivalence between the two forms.
    In the present case, involving light ions,  	it was observed that simultaneous calculations performed using the heavy-ion scheme (or similar variations as the default one from the \textsc{Fr2in} code \cite{BrownReactionCodes}) were sensibly different      in prior and post form,
         suggesting that the approximation itself may not be suitable.
 	 	The simultaneous calculations discussed here were thus performed as described above but keeping the reduced masses associated with the correct kinetic terms (e.g.~$K_{a\mu}$ and $K_{A\nu}$ for the core--$\mu$ and core--$\nu$ single-particle states for the projectile);      such modification almost perfectly restores the equivalence between the prior- and post-form Hamiltonians.      The geometry of the binding potentials was defined correspondingly: for instance, the single-particle state computed using kinetic energy $K_{A\nu}$ is associated      with a potential which attempts to account for the \nuclide{A}--\nuclide{\nu} interaction. 
                   Such choice also allows to adopt wave-functions with precisely the same form in both first- and second-order contributions to the transition amplitude ($\mathcal T^{(1)}$ and $\mathcal T^{(2)}$).
    However,      we notice that the precise geometry of binding potentials has only a limited impact on results, since any change is partly compensated for      by the      adjustment to the desired binding energy      (see the comment to \cref{eqSuperpositionXCoordinates}).
         The next step toward a fully consistent calculation would be to construct the three-particle state of $\mathcal{A}$, to be used in the first-order simultaneous calculation, as the superposition of states factorized in the $a$--$\mu$ and $A$--$\nu$ motions (“Y coordinates"), and similarly for $\mathcal{B}$.

    Finally, we point out that, for the very light nuclei of interest here, the contribution of the nucleon-nucleon potential $V_{\mu\nu}$ is not negligible compared to the other interactions (for instance, both the $\nuclide{\alpha}+\nuclide{p}$ and the $\nuclide{\alpha}+\nuclide{n}$ systems are unbound, even though the \nuclide[6]{Li} is bound).      Albeit not explicitly included, the impact of $V_{\mu\nu}$      is, in general, necessarily reabsorbed in an effective way in the other binding potentials ($V_{a\nu}$, for instance), since these must be      adjusted to fix the binding energy
    of the $a+\mu+\nu$ system           (and similarly for $\mathcal{B}$),
    as mentioned above.      In other words, {when $V_{\mu\nu}$ is neglected}, $V_{a\nu}$      cannot be the “true" $a$--$\nu$      interaction, but is a phenomenological potential.
    The downside is that such effective adjustment will then also appear in the transition operator
    of \cref{eqTwoNucTranHamiltExplicit},
    where it is not needed \cite[sec.~2]{Thompson2013}, potentially leading to an overestimation of the transfer cross-section for light nuclei,
        We expect this problem to be the main cause of
        the discrepancy observed between the results of 
        sec.~\ref{secTNTcrossection} and the experimental transfer cross section.
    We underline that the issue is a general feature of the aforementioned ``heavy-ion scheme'',      whose relevance however depends on the relative importance of the nucleon-nucleon interaction, $V_{\mu\nu}$, with respect to the core-nucleon binding potential.

\subsection{Numerical calculations}\label{secMethodologyNumerical}

         The numerical cross-section calculations shown in this work, both for one- and two-particle transfers, were carried out,      with the specifications given later      in sec.~\ref{secResults}, employing the \textsc{Fresco} code~\cite{Thompson1988} with the Numerov method.
         As the collision energies of interest are rather small, it has been checked that the transfer calculations are already well converged when truncating the partial-wave expansion of the scattering wave-functions at a total angular momentum of $7/2$.
    It is worth mentioning that the practical implementation in \textsc{Fresco} actually employs a coupled reaction channels (CRC) formulation, rather than the integral one in \cref{eqTAmplitudeOneparticleTransfer}, with an appropriate choice of the couplings to recover the DWBA approximation. The results can be shown to be equivalent in either formulation~\cite[sec.~3.4.2]{Satchler1983Direct} \cite{Ascuitto1969}.

    We also note that, in the (one- or two-particle) transfer calculation, the inclusion of      spin couplings (or, in general, an angular-momentum dependence) in the core-core potential is computationally very challenging and is not implemented in the \textsc{Fresco} code (or, to our knowledge, in any other available code, with few specific exceptions not covering the case here of interest, see e.g.~\ccite{GomezRamos2015}).
    Similarly, while spin-coupling terms in the projectile-target optical potentials can be fully taken into account for the construction of distorted-waves (the $\chi$ in \cref{eqTAmplitudeOneparticleTransfer}), the $U_{\mathcal A b}$ and $U_{a\mathcal B}$ terms explicitly appearing in \cref{eqTAmplitudeOneparticleTransfer} are truncated to their central, spin-independent parts.
    This implies that cross sections computed in prior and post form will not be equivalent if spin couplings in projectile-target potentials are included.
    Nevertheless,      such couplings allow a richer description of the system, 
         and were thus included in the present calculations, as detailed in sec.~\ref{secResults}.                All transfer results shown in this work were computed in post form (for first-order terms) and post-post form (for second-order terms): we deem this to be the choice yielding the best approximation of the correct result, given the form of the adopted potentials. 
    In particular, it is thought that the neglected non-central components in the projectile-target and core-core potential can better compensate with each other when they bear a more similar form 
    (each with respect to its own coordinates).
    The $\nuclide[6]{Li}+\nuclide{p}$ potential is the only one bearing a spin-spin component,
     thus the post form in the first (or single) transfer step is adopted to avoid neglecting it.
    Regarding the form for the second step of the {second-order} process,      we expect the non-central components of the core-core
    \nuclide{\alpha}--\nuclide{d} interaction to resemble more closely those of the \nuclide{\alpha}--\nuclide[3]{He} potential, rather than those of the \nuclide[5]{Li}--\nuclide{d} one.
    Additionally, the approximation involved in the post form for this step is more coherent with the analogous one made in the post-form simultaneous transfer (in both cases, the non-central part of the same \nuclide[3]{He}--\nuclide{\alpha} potential is being neglected), thus one may suppose that the coherent sum of both contributions will generate a more consistent result adopting the post form for all transfer steps (see \ccite[sec.~4.1.3.b, 4.2.3.e]{PerrottaTesiPhD} for more discussion).

\section{Results}\label{secResults}

\subsection{Deuteron transfer}\label{d results}
Considering the deuteron as a single  particle, the following optical potentials are needed 
to evaluate the transfer cross section:
 {$\nuclide[6]{Li} - \nuclide{p}$}, {$\nuclide{\alpha} - \nuclide[3]{He}$} (initial- and final-state projectile-target), and {$\nuclide{\alpha} - \nuclide{p}$} (core-core). All potentials employed in this work were parameterized as reported in section~\ref{secPotentialsDefinition}, and the values of the corresponding parameters are given in \cref{tabPotPar}.

For the core-core {\nuclide{\alpha} -- \nuclide{p}} interaction, we considered the potential quoted in~\ccite{Bang1979}, taking for the Coulomb part the potential generated by a uniformly charged sphere (the form in \cref{eqCoulombWoodsSaxonDefinition}) with radius fitted on elastic scattering experimental cross-sections at the relevant energies (see \ccites{Nurmela1998,Godinho2016} and references therein).
   
\begin{figure}[tbp]\centering\includegraphics[keepaspectratio = true, width=\linewidth]{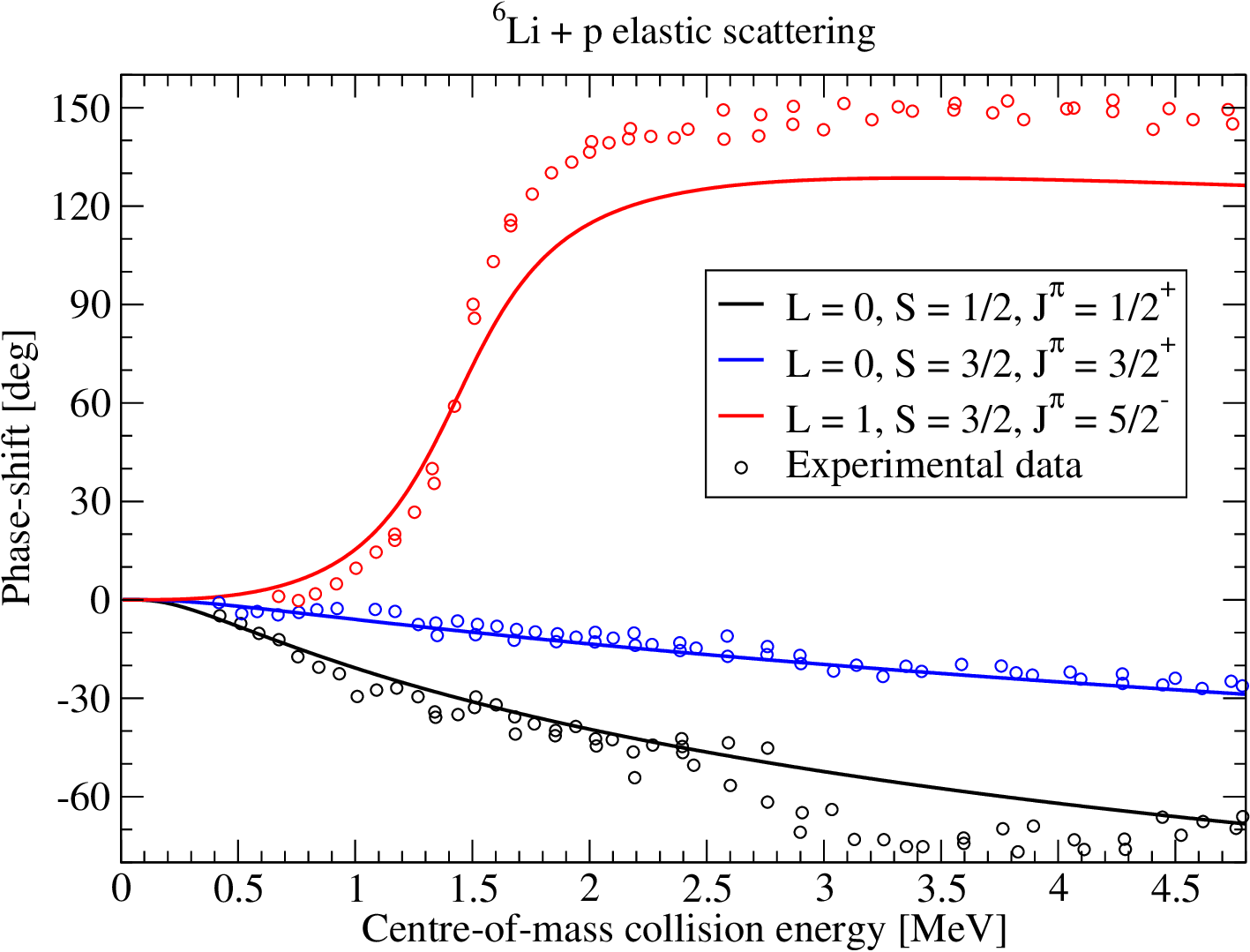}
\caption{\label{fig2}%
	Points are experimental $\nuclide[6]{Li}+\nuclide{p}$ elastic-scattering phase-shifts,  			collected in \ccite{Petitjean1969} from earlier experiments (see references therein). Lines in corresponding colors are the prediction of the optical potential discussed in sec.~\ref{d results}.  			 			In the figure legend, $\PTom$ is the $\nuclide[6]{Li}$--$\nuclide{p}$ relative orbital angular momentum modulus quantum number, $S$ the modulus quantum number for the sum of $\nuclide[6]{Li}$ and $\nuclide{p}$ spins, $\totAM$  			the modulus quantum number for the sum of $\vec \PTom$ and $\vec S$, and $\totP$ the parity of the state.}
	\end{figure}

Regarding the  \nuclide[6]{Li} -- \nuclide{p} potential, 
  most optical potentials in literature  include at most spin-orbit couplings. 
However, a term coupling the spins of both reactants is needed
 to reproduce the experimental phase-shifts. 
 Furthermore, to our knowledge, no published energy-independent potential compares acceptably with the elastic scattering cross-section in the energy range of interest here.
	For this work, a real energy-independent  	potential was adjusted to reproduce the most relevant partial waves of the experimental phase-shifts, namely the $s$-waves and the $p_{5/2}$ wave, the latter showing a resonant trend,  	attributed  	to a \nuclide[7]{Be} level, which is visible also in the transfer channel.
	The result is shown in \cref{fig2}.  	The potential found in this manner introduces no spurious resonances (namely, resonances with incorrect energy or quantum numbers) in the region of interest.  An imaginary component was subsequently added to the interaction to improve the agreement with the elastic scattering experimental cross-sections below \qty{1.4}{\MeV} reported in~\ccites{McCray1963,Dubovichenko2011}.

For the {$\nuclide{\alpha} - \nuclide[3]{He}$} optical
potential, instead, the real and imaginary parts were fitted to the elastic scattering cross-section 
data presented in refs.~\cite{Tombrello1963,Spiger1967}, using \textsc{Sfresco}~\cite{Thompson1988}.

The remaining ingredients are the $\Braket{\nuclide{\alpha} \, \nuclide{d} | \nuclide[6]{Li} }$  and $\Braket{ \nuclide{p} \, \nuclide{d} | \nuclide[3]{He} }$ overlap functions.
In each case, the asymptotic radial trend of each wave-function is fixed by adjusting the associated binding energy to the correct separation energy.  The $\nuclide{\alpha}+\nuclide{d}$ wave-function is constructed with the same procedure adopted in \ccite[sec.~5.3]{Nishioka1984}, but using different choices for the relative weight of $s$- and $d$-wave components, as detailed  in sec.~\ref{secDeuteronTransferRoleDeformation}.   Note that, according to this construction, the $s$ and $d$-wave radial wave-functions have, respectively, one and zero nodes,  as in ref.~\cite{Nishioka1984} and as suggested by the Wildermuth connection \cite[eq.~(16.32)]{Satchler1983Direct}), while other works  suggest one node for the $d$-wave a well  \cite[sec.~V.C]{Forest1996}, \cite[sec.~IV]{Lehman1982}.
   An overall spectroscopic factor of \num{0.82} is then assigned to the wave function, such that the $s$-wave state reproduces the asymptotic normalisation coefficient of \SI[per-mode=power]{2.29}{\femto\metre\tothe{-1/2}} quoted in \cite[sec.~4.2]{Mukhamedzhanov2022}; this is also the spectroscopic factor found in \cite[sec.~V.C]{Forest1996} for only the $s$-wave component.
 
The $\Braket{ \nuclide{p} \, \nuclide{d} | \nuclide[3]{He} }$ overlap function was constructed using the binding potential reported in \cite[tab.~VIII]{Brida2011}, which reproduces a Green's Function Monte Carlo (GFMC) overlap  while providing the correct asymptotic form for the wave-function.
 The state includes both an $s$ and $d$-wave component, with no nodes in the radial parts,  and with spectroscopic factors of \num{1.31} and \num{0.0221} respectively \cite[tab.~IV]{Brida2011}.
 Their sum is  slightly smaller than the spectroscopic factor  found within the independent-particle
Shell Model, $3/2$,  and is  consistent with values  used in previous (\nuclide[3]{He},\nuclide{p}) or (\nuclide{p},\nuclide[3]{He}) calculations~\cite{Werby1973,Ayyad2017}.

\begin{figure}[tbp]
    \centering
    \includegraphics[keepaspectratio=true, width=\linewidth]{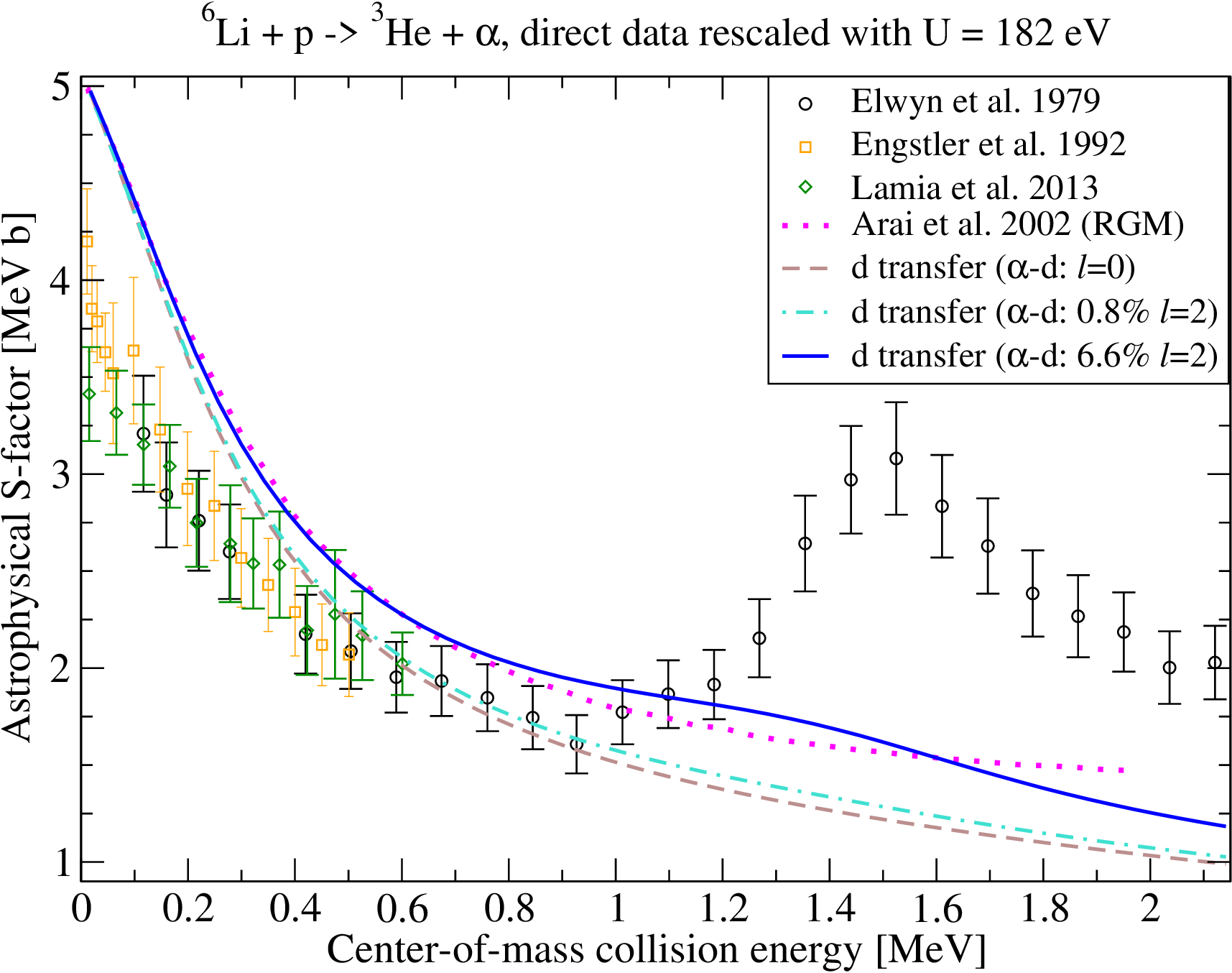}
    \caption{\nuclide[6]{Li}(\nuclide{p},\nuclide[3]{He})\nuclide[4]{He} astrophysical $S$-factor. Points are a subset of the rescaled data in the bottom panel of \cref{fig1} (only some datapoints are shown for readability).
    The magenta dotted line is the resonating-group-method calculation in \ccite{Arai2002}. The other lines are the present-work deuteron-transfer post-form DWBA calculation.    
         The brown dashed line was computed including only a relative orbital angular momentum $\ell=0$ (“$l$" in the figure legend)
    for the \nuclide[4]{He}--\nuclide{d} wave-function. The turquoise dot-dashed and blue solid line were computed including also a $d$-wave component (constructed as reported in sec.~\ref{d results}) with a relative norm of about \qty{0.8}{\percent} and \qty{6.6}{\percent} respectively.}      \label{fig:dtransfer}
\end{figure}

The astrophysical factors obtained in DWBA
 with the above ingredients
are displayed in \cref{fig:dtransfer}.  The region of the resonance, 
which corresponds to the second $\frac{5}{2}^-$ state of \nuclide[7]{Be}, is clearly underestimated with respect to the data:  this is not surprising, considering that the optical potential in the exit channel was fitted on global features of the elastic cross section at smaller energies.   The limited number of parameters considered to describe the optical potentials in the entrance and the exit channels makes it difficult to fit the whole energy dependence of the cross section. Our aim here is to focus on the trend at energies below \qty{1}{\MeV}.   

     It is interesting to notice that the calculations
can reproduce the almost linear trend  observed in the energy range  of astrophysical interest,
where       our results even overestimate the bare cross-sections extracted from experimental data.      Hence, the single-particle transfer scenario, which implicitly assumes a clustered deuteron structure, would seem to support the additional (“anomalous") enhancement observed in the direct data.
 A similar result  was obtained from
 the Faddeev three-body calculation in \ccite{Vasilevsky2009},
and the resonating-group method (RGM) calculations of  \ccites{Arai2002} (the magenta dotted line in \cref{fig:dtransfer}) and~\cite{Solovyev2018}.
  In particular, we also note the quite good agreement between the present  calculation with the largest $d$-wave contribution (blue solid line {in \cref{fig:dtransfer}}) and the RGM
result  from \ccite{Arai2002}.

\subsubsection{Role of deformations in the inter-cluster wave-functions}\label{secDeuteronTransferRoleDeformation}

\begin{figure}[htbp]\centering\includegraphics[keepaspectratio=true, width=0.7\linewidth]{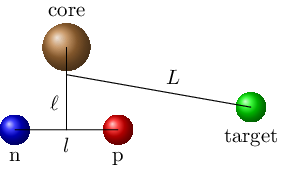}
\caption{\label{figCoordinate}%
Representation of the reactants structure and the Jacobi “$T$" coordinate system. Spheres depict the particles treated as elementary in the present formalism (in sec.~\ref{d results}, the transferred \npPair{} system is inert). Each line represents a Jacobi “$T$" coordinate (transferred system internal motion, core-transferred motion, projectile-target motion), and $\nnom$, $\ctom$, $\PTom$ are the symbols employed here to represent the orbital angular momentum modulus quantum number associated with each of those coordinates.}
\end{figure}

Here, we would like to address the impact of possible deformations effects, in the reactants ground state, on the low-energy trend of the astrophysical factor. Indeed, the Coulomb repulsion could be reduced
 if deformation effects are taken into account in the \nuclide[6]{Li} cluster-model ground state,
thus providing a further explanation of the cross section enhancement associated with the electron screening problem \cite{Spitaleri2016}.
       In the single-particle transfer calculations,  clustering is being imposed in the sense that \nuclide{d} and \nuclide{\alpha} are treated as point-like particles, and the $\Braket{\nuclide{\alpha} \, \nuclide{d} | \nuclide[6]{Li} }$ ground-state overlap function is a two-body \nuclide{\alpha}--\nuclide{d} wave-function (and analogously for \nuclide[3]{He}).  Usually, the \nuclide[6]{Li} two-cluster-model ground state is associated with a wave-function with \nuclide{\alpha}--\nuclide{d}  relative angular momentum, $\ctom_{\nuclide{\alpha}\nuclide{d}}$, fixed to 0 (see e.g.~\cite{Kubo1972,Muskat1995,DiazTorres2003,Woodard2012,Zagatto2022}):  the brown dashed line in \cref{fig:dtransfer} is a calculation performed in this manner (including only the $s$-wave component).
    However, a small $d$-wave ($\ctom_{\nuclide{\alpha}\nuclide{d}}=2$) component is suggested by models and required to reproduce some experimental structure properties of \nuclide[6]{Li} \cite{Nishioka1984,Merchant1985,PerrottaSantaTecla2022}.   To take this into account, the calculation represented by the turquoise line in \cref{fig:dtransfer} also  includes an \nuclide{\alpha}--\nuclide{d} $d$-wave component with a relative norm of \SI{0.83}{\percent} (keeping the total norm of the bound state fixed): this was  adjusted to reproduce the \nuclide[6]{Li} electric quadrupole moment (as in \ccite[sec.~5.3]{Nishioka1984} but using the updated experimental values in \ccite{IAEAElectroMagneticMoments}).
Finally, the blue solid line corresponds to a calculation with an \nuclide{\alpha}--\nuclide{d} wave-function reproducing the experimental \nuclide[6]{Li} magnetic dipole moment (see \ccites[sec.~5.2.1]{PerrottaTesiPhD} and \cite{PerrottaSIF2021} for more details), namely with a relative $d$-wave norm of \qty{6.6}{\percent}.

The deformed components of the inter-cluster wave-functions allow the transfer to proceed  through partial waves which would be otherwise forbidden. This is because the interactions adopted in the present calculation (including only central terms and vector spin-orbit  couplings) 
 conserve not only the total angular momentum and parity of the system, $\totAM^\totP$, but also the modulus of the total orbital angular momentum of the system, which in each partition can be expressed as the sum of $\ctom$ (defined above)  and the orbital angular momentum between the projectile and target centers of mass,  $\PTom$, see \cref{figCoordinate}. 
It is stressed that the projectile-target coordinate is different in the initial and final partitions, thus $\PTom_{\nuclide{Li}\nuclide{p}}$ and $\PTom_{\nuclide{\alpha}\nuclide{He}}$  are distinct quantities.
However, if both the {$\Braket{\nuclide{\alpha} \, \nuclide{d} | \nuclide[6]{Li} }$}  and {$\Braket{ \nuclide{p} \, \nuclide{d} | \nuclide[3]{He} }$}  overlap functions include only an $\ctom=0$ component,  then it must be $\PTom_{\nuclide{Li}\nuclide{p}} = \PTom_{\nuclide{\alpha}\nuclide{He}}$.   For instance, a $\nuclide[6]{Li}+\nuclide{p}$ pair colliding in $s_{3/2}$ wave ($\totAM^\totP = \frac{3}{2}^+$ and $\PTom_{\nuclide{Li}\nuclide{p}} = 0$) cannot couple to the $\nuclide[3]{He}+\nuclide{\alpha}$ channel, where $\totAM^\totP = \frac{3}{2}^+$ would require $\PTom_{\nuclide{\alpha}\nuclide{He}} = 2$. The same reasoning forbids an incoming $p_{5/2}$ wave ($\PTom_{\nuclide{Li}\nuclide{p}} = 1$) if $\ctom\neq0$ components are discarded.
It is worth noting that the
{$\nuclide[6]{Li}+\nuclide{p}$ $p_{5/2}$ is the wave displaying a resonance in the elastic scattering phase-shift}
fitted in our calculations (see \cref{fig2}), which has been associated with the \nuclide[7]{Be} $\frac{5}{2}^-$ excited state at about \qty{7.2}{\MeV}~\cite{Tilley2002} and to the aforementioned peak in the transfer channel~\cite{Tumino2003}.
As a consequence, the inclusion of non-spherical overlap functions is generally expected to alter the reaction cross-section: if the total spectroscopic factor of the state is kept constant, the cross-section will increase or decrease depending on the relative probability of the processes allowed by the sperical and non-spherical components.

If the adopted initial- and final-state projectile-target interactions do not include spin-coupling terms,  the difference induced by the
inclusion of the $d$-wave component of the structure wave functions would depend just on the number of spin states $J^\pi$ associated with the partial waves allowed by each wave-function component, and on the distorted waves associated with such partial waves (in general, greater projectile-target angular momenta are disfavored). In a test calculation where the aforementioned spin couplings were removed,
we found that, for collision energies below about \qty{1}{\MeV}, the variation of the astrophysical $S$-factor  induced by the deformed component of the wave-functions is almost constant with energy.    However, the presence of spin-coupling terms in the optical potentials
(for instance the \nuclide[6]{Li}--\nuclide{p} spin-spin term mentioned above),   which are included 
in our calculations,
can change this conclusion,
as detailed in the following.  
                    
   As shown in \cref{fig:dtransfer}, the transfer $S$-factor increases at energies around the Coulomb barrier when deformed components are included in the overlap functions (see, in particular, the solid blue line).
A partial-wave decomposition of the computed cross-sections reveals that,  for collision energies above about \qty{1}{\MeV}, the difference between the brown dashed line in \cref{fig:dtransfer} and the other two calculations is mainly due to a contribution with $\nuclide[6]{Li}+\nuclide{p}$ incoming in $p_{5/2}$ wave (the resonant wave, see above), which leads to an enhancement at a slightly smaller energy with respect to  the peak observed in the experimental transfer data.
            We underline again that the potentials employed in this work were more focused on describing the region at sub-Coulomb collision energies.
Moreover, the coupling to excited states of \nuclide[6]{Li} is expected to play a non-negligible role in reproducing the resonant trend (note that the \nuclide[6]{Li} breakup channel opens at a center-of-mass collision energy of about \qty{1474}{\keV}).  The inclusion of deformation effects can thus be of interest for future investigations of the region around the barrier using coupled-reaction-channels approaches,  since standard  studies typically assign states with definite inter-cluster orbital angular momentum modulus to each level  (see e.g.~\ccites{Muskat1995,DiazTorres2003,Woodard2012,Zagatto2022}).

Regarding the region of astrophysical interest, it is interesting to see that  the cross-section difference between the calculations involving spherical or deformed overlap functions in \cref{fig:dtransfer} vanishes at
very low collision energies.
Due to the adopted spin-coupling terms in the optical potentials, we observe that the increase in the $s_{3/2}$-wave cross-section  equals the reduction in the $s_{1/2}$-wave cross-section, thus leading to a negligible total variation.               Therefore, within the present DWBA framework,  we conclude that 
possible static deformation effects, associated with clustered configurations, do not play a role in the observation of an abnormal electron screening, though they affect the overall trend of the astrophysical factor. 
This is also in agreement with the findings in \cite{PerrottaSIF2021} regarding the \nuclide[6]{Li}--\nuclide{p} barrier penetrability.
  
The discussion of dynamical deformation effects, namely polarization or reorientation effects, 
requires to go beyond the present DWBA description of the reaction,  as we plan to do in future works.

\subsection{Two-nucleon transfer}
\label{np results}

In this section we discuss the results obtained with second-order DWBA calculations. 
The ingredients of the two-nucleon transfer process which are in common with the deuteron-transfer calculation  (see section~\ref{d results}) were chosen to be the same.  This is in particular the case for the {$\nuclide[6]{Li} - \nuclide{p}$}, {$\nuclide{\alpha} - \nuclide[3]{He}$}, and {$\nuclide{\alpha} - \nuclide{p}$} potentials.  The role of the intermediate partition in second-order DWBA requires to define more optical and binding interactions.  The {\nuclide{\alpha}--\nuclide{d}} core-core potential was, for consistency, taken to be the same one employed in section~\ref{d results} to construct {each component} of the \nuclide[6]{Li} bound state, but with the depth of the volume term rescaled to match the potential in \ccite{Gammel1960} at zero distance (whose numerical value is given in \ccite[fig.~2]{Kubo1972}).
 The fact that both $^5$Li and $^5$He are unbound makes it difficult 
 to fit the interactions involving these systems. 
Thus, we rely on generic  optical potentials, namely the one in~\ccite{Daehnick1980} for the {\nuclide[5]{Li}--\nuclide{d}} projectile-target interaction,  and the one in \ccite{Varner1991} for the {\nuclide[5]{Li}--\nuclide{p}} core-core interaction. The form and parameters of the adopted potentials are reported in Appendix~\ref{secPotentialsDefinition}.
 
\subsubsection{Description of the ground-state configuration}\label{secThreeparticlewavefunctions}  
The new important aspect, with respect to the calculations in sec.~\ref{d results},  is that the $\Braket{\nuclide{\alpha} \, \nuclide{p} \, \nuclide{n} | \nuclide[6]{Li} }$  and $\Braket{ \nuclide{p} \, \nuclide{n} \, \nuclide{p} | \nuclide[3]{He} }$  overlap functions are three-body wave-functions. The transferred system is thus not fixed to an inert deuteron anymore, and it is possible to  implement different degrees of deuteron clustering within the composite system.   The specific approach  employed for the construction of the wave-functions  is discussed in greater detail in \ccite{PerrottaTesiPhD}, and summarized in the following.
 
The three-particle wave-functions  are expressed using \cref{eqSuperpositionXCoordinates} as described in sec.~\ref{secMethodTNTSimultaneous}.
To ensure that a complete state $\Psi$ has the correct binding energy, the binding energy of each single component of the superposition, for instance $\phi_{a\mu,i} \phi_{a\nu,i}$ in \cref{eqSuperpositionXCoordinates}, is fixed to the same value.  Note that this, in general, implies that the binding potentials will be different for each component $i$.
The standard prescription (see e.g.~the construction implemented in the \textsc{Fr2in} code \cite{BrownReactionCodes})  is to assign to each single-particle state, $\phi$, a binding energy equal to half the total separation energy of $\mathcal A$ (or $\mathcal B$). Such prescription was adopted here for the \nuclide[6]{Li} state (where the physical \nuclide{\alpha}--nucleon states are unbound). Following the approach discussed in sec.~\ref{secMethodTNTSimultaneous},  the core-proton wave-functions are constructed using the reduced mass of the \nuclide{\alpha}--\nuclide{p} system, while the core-neutron wave-functions involve the reduced mass of the \nuclide[5]{Li}--\nuclide{n} system. The binding potential employed to construct these wave-functions is the same \nuclide{\alpha} -- \nuclide{p} interaction discussed in sec.~\ref{d results} and employed as core-core potential, with the following differences. First, for each single-particle wave-function the volume depth is adjusted to reproduce the aforementioned binding energy. Second, for the core-neutron wave-functions, the Coulomb term is removed, and all radii were rescaled by $(6/5)^{1/3}$ to empirically account for the different size of \nuclide[5]{Li} with respect to the \nuclide{\alpha} (this was seen to yield negligible differences in the results.    The components of the superposition defining the \nuclide[6]{Li} ground state  were chosen following the results of the Faddeev 3-body calculation in~\ccite{Bang1979} (which adopts the same \nuclide{\alpha}--nucleon interaction in use here). The paper reports  the weights of the components of the computed wave-function   in a different angular momentum coupling scheme than the one adopted here (which is the “$jj$" scheme), thus they were transformed accordingly.
Some of the components quoted in \ccite[tab.~2]{Bang1979} for the \nuclide[6]{Li} ground state are not compatible with the present model and calculation scheme (for instance, the configuration with total isospin 1) 
 and were thus discarded.
The adopted components comprise a total norm of 0.936 and are shown in \cref{tab6LiThreeParticle}.
  \begin{table}[tbp]
    \caption{\label{tab6LiThreeParticle}%
    Decomposition of the $\nuclide{\alpha} + \nuclide{p} + \nuclide{n}$ three-particle wave-function adopted for the \nuclide[6]{Li} ground state (see top left panel in \cref{fig6LiTNTpdf}), deduced as detailed in the text.
        The state is decomposed into products of single-particle states in ``$jj$'' angular momentum coupling scheme, as specified in the first two columns, marking respectively the proton and neutron state (in the present notation, the radial quantum number corresponds to the number of radial nodes plus one). The last column is the amplitude associated with each component.}
    \centering
    \begin{tabular}{ccS[table-format=+1.4]}         \nuclide{p} shell & \nuclide{n} shell & {Amplitude} \\ \toprule
                 $1p_{3/2}$	& $1p_{3/2}$	&  0.7482	\\
        $1p_{3/2}$	& $1p_{1/2}$	& -0.4044	\\
        $1p_{1/2}$	& $1p_{3/2}$	&  0.4044	\\
        $1p_{1/2}$	& $1p_{1/2}$	& -0.1228	\\
        $2s_{1/2}$	& $2s_{1/2}$	& -0.1843	
    \end{tabular}
\end{table}
  The data included in \ccite{Bang1979} is not sufficient to extract all relative phases between the different components in a straightforward manner. To find a set of reasonable relative phases, the weight of each component
was compared with the one given by a three-body calculation in Hyperspherical
Harmonics formalism \cite{Casal2021Private}. 
 With the choice of phases in \cref{tab6LiThreeParticle} for the components in \ccite{Bang1979}, the two calculations predict
a similar structure for the global wave function and comparable norms for the different components in “$jj$" scheme.

Regarding instead the \nuclide[3]{He},
as is customary (consider for instance  the construction implemented in the \textsc{Fr2in} code \cite{BrownReactionCodes}), the total wave-function for \nuclide[3]{He} comprises only one  component, in which  both single-particle wave-functions are $1s$ states (i.e.~with  no nodes in the radial part).   However,  following the approach discussed in sec.~\ref{secMethodTNTSimultaneous},
in this work the core-neutron wave-function is  constructed using the \npSystem{} reduced mass, the $\nuclide{p}+\nuclide{n} \to \nuclide{d}$ experimental separation energy, and a \npSystem{} binding potential from the \textsc{Fr2in} code \cite{BrownReactionCodes}, while the core-proton wave-function is the $1s$ component of the \nuclide{d}--\nuclide{p} wave-function discussed in sec.~\ref{d results} for the deuteron-transfer calculation.

\begin{figure}[tbp]\centering\includegraphics[keepaspectratio=true, width=\linewidth]{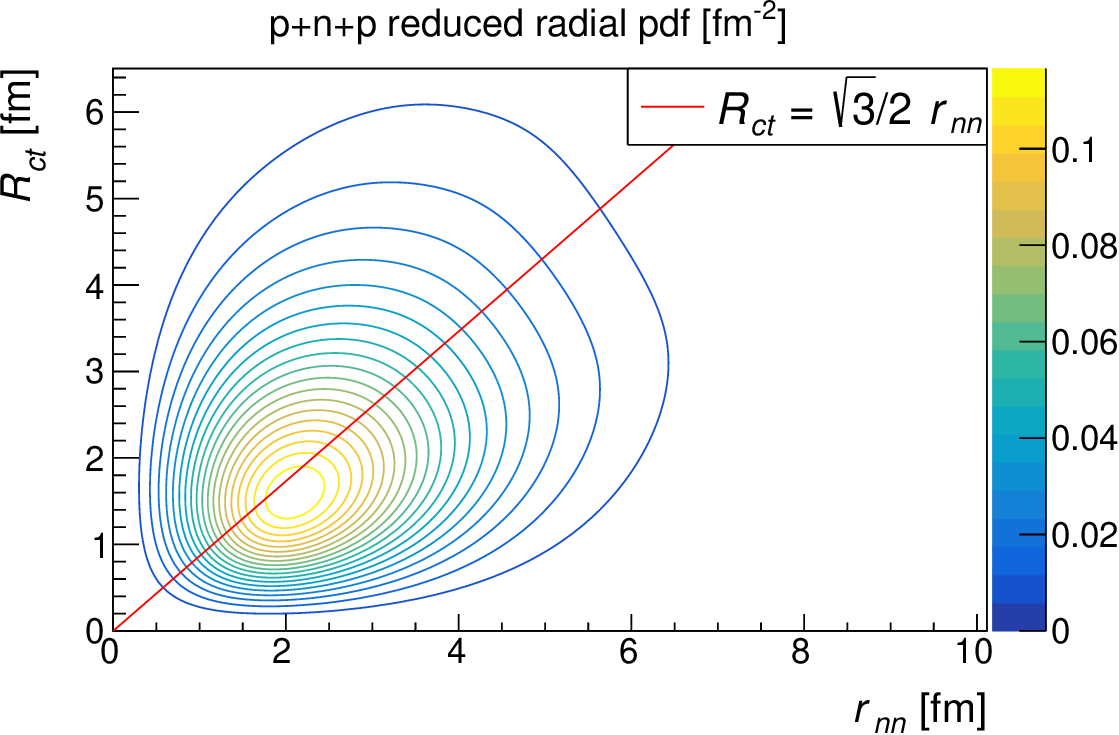}%
\caption{\label{fig3HeTNTpdf}%
Contour plot for the reduced radial probability density function (“pdf”), in \si{\per\femto\metre\squared}, for the core + \nuclide{p} + \nuclide{n} state of \nuclide[3]{He}. The $x$-axis is the distance between the transferred \nuclide{p} and \nuclide{n}, while the $y$-axis is the distance between the core and the \npSystem{} center of mass. The line obeys  		the equation in the legend. See sec.~\ref{secThreeparticlewavefunctions} for details.}
\end{figure}

\begin{figure*}\centering\includegraphics[keepaspectratio=true, width=0.49\textwidth]{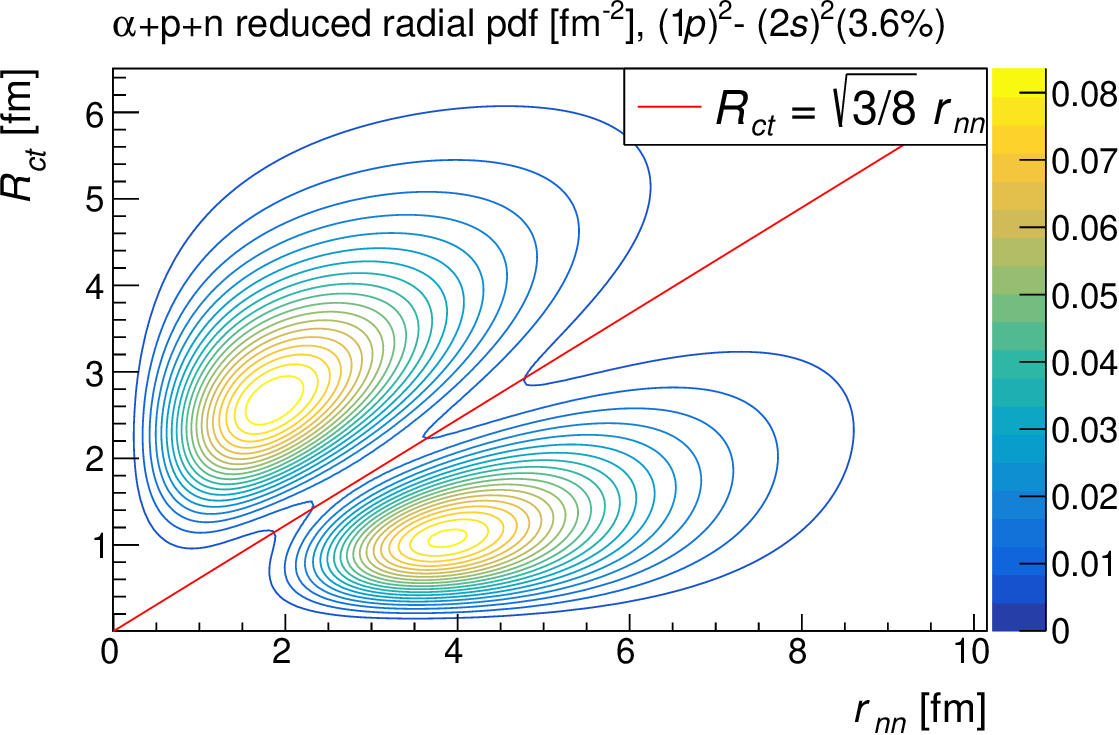}\hfill
\includegraphics[keepaspectratio=true, width=0.49\textwidth]{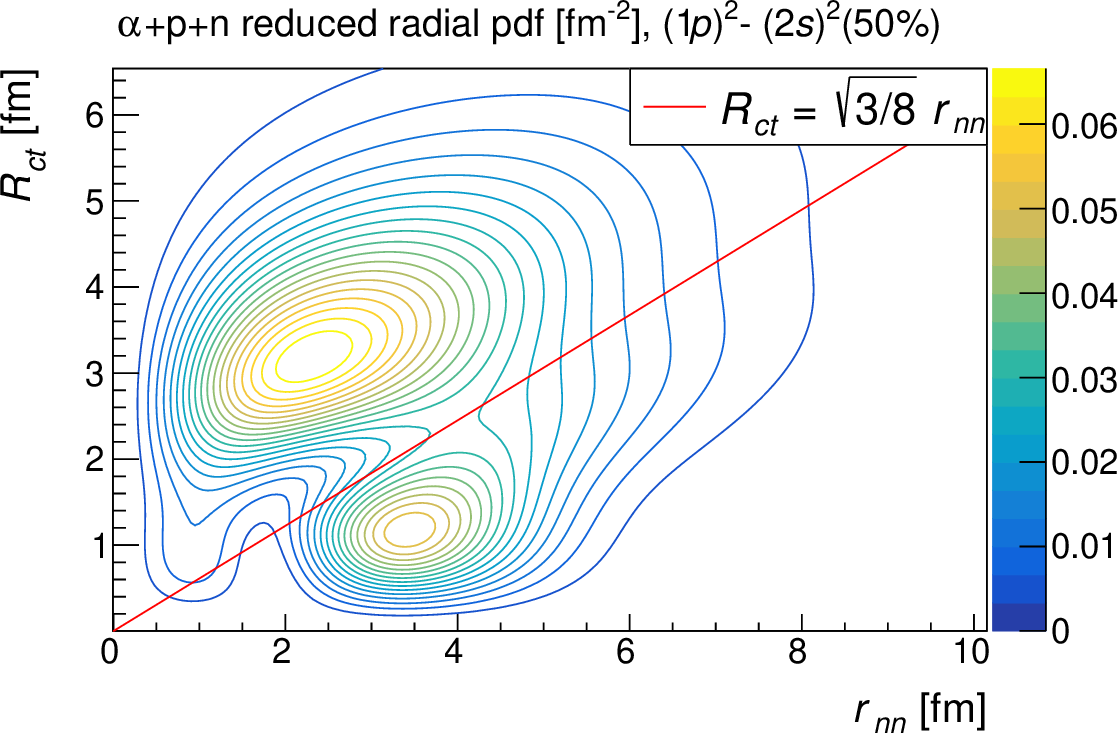}\newline
\includegraphics[keepaspectratio=true, width=0.49\textwidth]{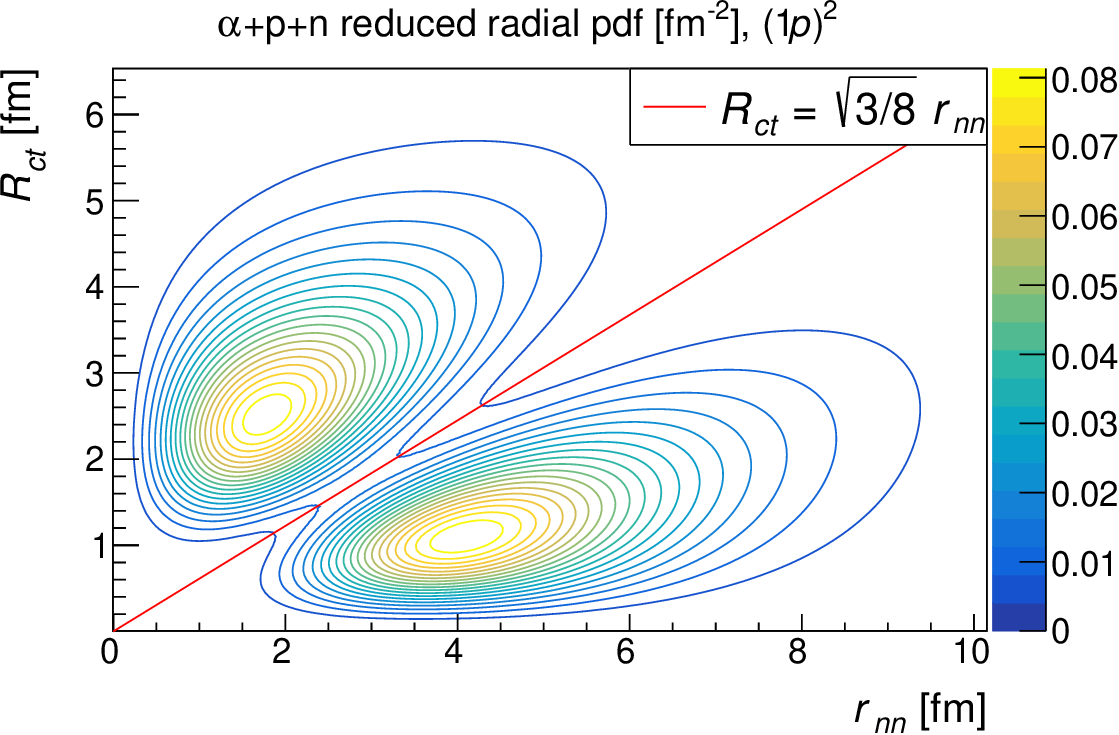}\hfill
\includegraphics[keepaspectratio=true, width=0.49\textwidth]{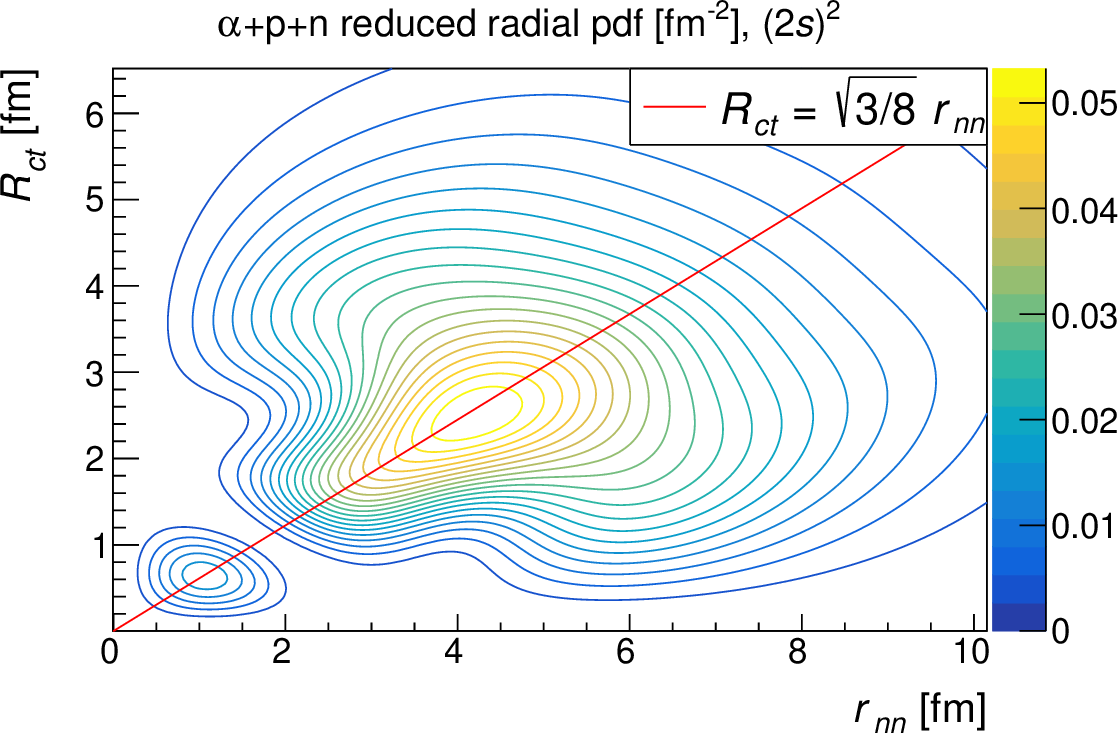}\newline
\includegraphics[keepaspectratio=true, width=0.49\textwidth]{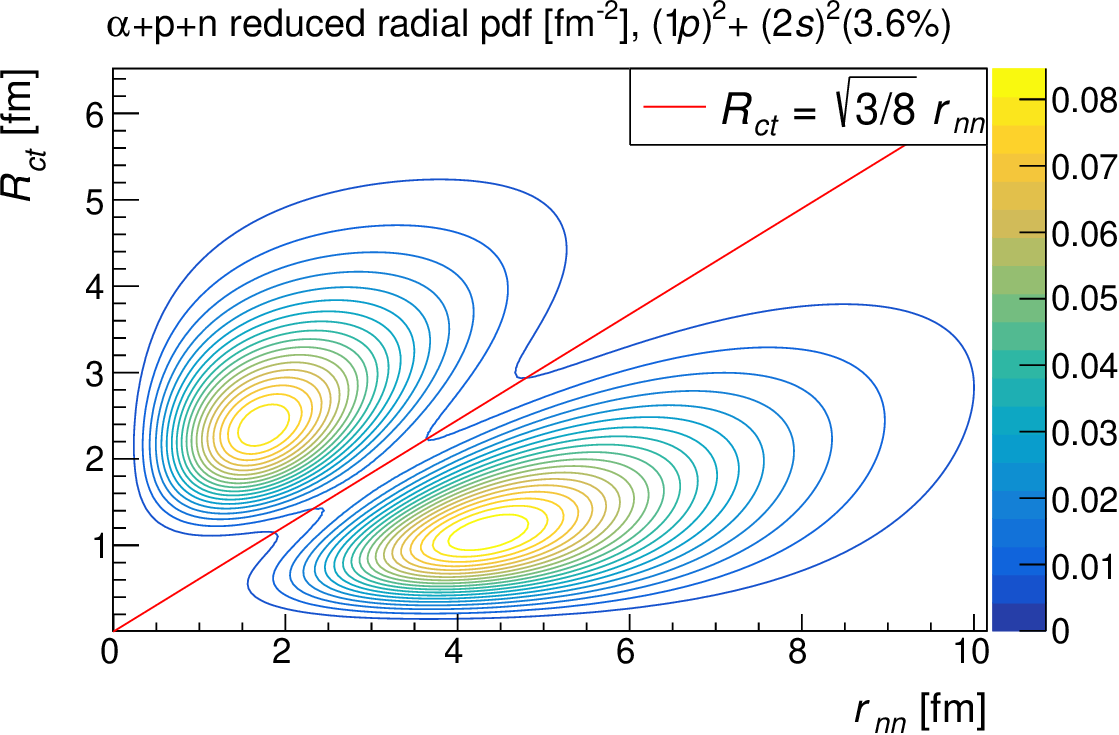}\hfill
\includegraphics[keepaspectratio=true, width=0.49\textwidth]{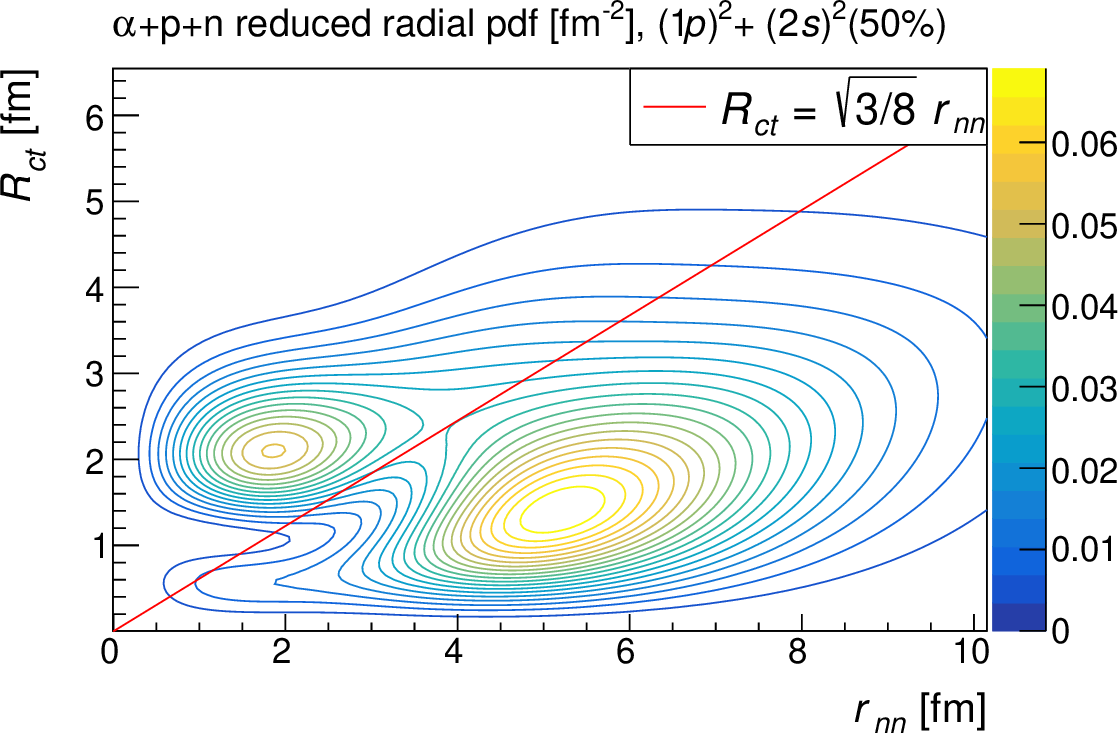}
\caption{\label{fig6LiTNTpdf}%
Same as \cref{fig3HeTNTpdf} but for \nuclide[6]{Li}.
        The top-left panel represents the state yielding the best agreement with available information from three-body calculations, constructed using the {amplitudes} in \cref{tab6LiThreeParticle}.          The state represented in the bottom-left panel was constructed with the same weights but with opposite sign for the “$2s_{1/2} \times 2s_{1/2}$” component {in \cref{tab6LiThreeParticle}.}          The states represented in the top- and bottom-right panels are constructed as those in the top- and bottom-left panels but assigning equal absolute weight to the $(2s)^2$ and $(1p)^2$ configurations.          The states shown in the middle-left and right panels do not include the configurations with transferred nucleons occupying, respectively, the \nuclide[6]{Li} $2s$ or $1p$ shells.                   See sec.~\ref{secThreeparticlewavefunctions} for details.}
\end{figure*}

These wave-functions are then translated,  using the Moshinsky coordinate transformation~\cite{Moshinsky1959},  into the so-called “T” Jacobi coordinates, namely, they are expressed as a function of the displacement between the transferred neutron and proton ($\vec r_{nn}$) and the displacement between the center of mass of the transferred system  and the rest of the nucleus, i.e.~the core ($\vec R_{ct}$).
 The Moshinsky-transformed wave-functions are finally employed for the simultaneous transfer calculation. The second-order contributions are instead computed using directly the form of the aforementioned single-particle wave-functions, $\phi$, for each transfer step (thus employing them in the so-called “Y" Jacobi coordinate system), weighted in the same way as the components for the three-particle wave-functions discussed above.
Note that this implies that a  bound wave-function is assumed for the intermediate \nuclide[5]{Li} state in the two-step transfer: the adopted $Q$-values for each transfer step are adjusted consistently.  A more accurate treatment might be sought through a description of \nuclide[5]{Li} continuum.  For both the first- and second-order calculations, the same overall spectroscopic factors employed in the deuteron-transfer case are adopted (see sec.~\ref{d results}): these spectroscopic factors are not included in \cref{fig3HeTNTpdf,fig6LiTNTpdf,tab6LiThreeParticle,tabClusteredPeak}.

\Cref{fig3HeTNTpdf,fig6LiTNTpdf} represent the probability density functions for the 
 ground state structure of \nuclide[3]{He} and \nuclide[6]{Li} respectively, in terms of the radial “T” Jacobi coordinates.  In both figures, the red line corresponds to the quadrant bisector {($x = y$)} in the rescaled Jacobi space coordinates, $x = \sqrt{A_{nn}} r_{nn}$ and $y = \sqrt{A_{ct}} R_{ct}$, where $A_{ab}$ is the reduced mass number for the $a$--$b$ system
(in particular, for the
$\nuclide{p}+\nuclide{n}+\nuclide{p}$ nucleus it is $A_{nn} = 1/2$ and $A_{ct} = 2/3$, whereas for the
$\nuclide{\alpha}+\nuclide{p}+\nuclide{n}$ nucleus it is $A_{nn} = 1/2$ and $A_{ct} = 4/3$).  The kinetic energy part of the Hamiltonian is symmetric under the exchange of the rescaled coordinates $x$ and $y$, thus explaining the nearly symmetric trend observed  for the probability density functions with respect to the $x = y$ 
line. 
  
The $R_{ct}= \frac{\sqrt{3}}{2} r_{nn}$ line, which is the $x=y$ line for \nuclide[3]{He}, corresponds to equilateral triangle configurations.
   Since the nucleons in \nuclide[3]{He} are placed in the $s$ shell, such line hosts
the maximum probability density for such system.
   
A completely different situation is found for 
$^6$Li, as expected. 
The choice of relative phases for the different components (see the discussion above) yielding the best agreement between the calculations in \ccites{Bang1979,Casal2021Private}  corresponds to the probability density function shown
in the top left panel of \cref{fig6LiTNTpdf}.
We observe two maxima at the opposite sides of the $x = y$
 line,  which corresponds to $R_{ct}= \sqrt{\frac{3}{8}} r_{nn}$ for $\nuclide{\alpha}+\nuclide{n}+\nuclide{p}$.
The peak at $R_{ct} > \sqrt{\frac{3}{8}} r_{nn}$, featuring a greater maximum probability density,  can be associated with a configuration in which the neutron and the proton are close together in space, forming a cluster resembling a deuteron, with a larger separation from the rest of the nucleus.  The other peak corresponds to a cigar-like configuration in which neutron and proton are far apart from each other 
and their center of mass is close to the $\alpha$ core. It can be useful to note that the \nuclide{\alpha} particle and the deuteron have a root-mean-square charge radius of about \qty{1.68}{\femto\metre} and \qty{2.14}{\femto\metre} respectively \cite{Angeli2013}.

For comparison, we have taken into account some alternative choices for the \nuclide[6]{Li} wave-functions.  As illustrated in the bottom panels of \cref{fig6LiTNTpdf}, the cigar-like configuration becomes more important by inverting  the relative phase between the $(2s)^2$ and $(1p)^2$ components.
An intermediate situation is found
if the $(2s)^2$ components of the wave-function are neglected, as shown in the middle left panel of \cref{fig6LiTNTpdf}.  The right panels of \cref{fig6LiTNTpdf} include additional test cases ({see Table II}) analogous to those in the left panels but assuming a greater norm for the $(2s)^2$ component, to enhance (or remove) the effect of the configuration mixing.
   \begin{table}[tbp]
		\caption{\label{tabClusteredPeak}
            For each tested \nuclide[6]{Li} wave-function, identified by the relative weight and phase of the adopted components in core-nucleon coordinates, the table lists, for only the “clustered region" $R_{ct} > \sqrt{\frac{3}{8}} r_{nn}$ (the region above the red line in \cref{fig6LiTNTpdf}, see sec.~\ref{secThreeparticlewavefunctions} for details):
            the norm under such region, normalized to the average of the total norm of the transformed wave-functions associated with the pure $(1p)^2$ and $(2s)^2$ configurations;
            the maximum of the radial probability density function within such region, normalized to the value for the pure $(1p)^2$ case.}                                 		\centering
		\begin{tabular}{cS[table-format=1.3]S[table-format=1.2]}    & \multicolumn{2}{c}{Clustered region} \\\cline{2-3}
\nuclide[6]{Li} PDF & {Integral} & {Maximum} \\ \toprule  $(1p)^2(\qty{50}{\percent})-(2s)^2(\qty{50}{\percent})$	& 0.711 & 0.80 \\
$(1p)^2(\qty{87.6}{\percent})-(2s)^2(\qty{12.4}{\percent})$	& 0.654 & 0.97 \\
$(1p)^2(\qty{96.4}{\percent})-(2s)^2(\qty{3.6}{\percent})$ & 0.596 & 1.00 \\
$(1p)^2(\qty{100}{\percent})$ & 0.516 & 1 \\
$(2s)^2(\qty{100}{\percent})$ & 0.468 & {--} \\
$(1p)^2(\qty{96.4}{\percent})+(2s)^2(\qty{3.6}{\percent})$ & 0.432 & 0.96 \\
$(1p)^2(\qty{87.6}{\percent})+(2s)^2(\qty{12.4}{\percent})$	& 0.365 & 0.89 \\
$(1p)^2(\qty{50}{\percent})+(2s)^2(\qty{50}{\percent})$	& 0.273 & 0.63
         		\end{tabular}
\end{table}
 \Cref{tabClusteredPeak} reports, for several cases,    the norm of the wave-function in the “clustered” region above the red line in \cref{fig6LiTNTpdf}.
The differences between the norms in each case are sizable,
  and seem to be related also to the
extension of each peak (in particular, how important each configuration is in peripheral regions).   Such differences can be expected to be relevant for the transfer process.

Qualitatively, our findings are similar to those in \cite{Catara1984}, where specific structures in the probability density functions are enhanced by mixing configurations in which nucleons lie in shells with different parity.

\subsubsection{Transfer cross-section}\label{secTNTcrossection}
 \begin{figure}[tbp]
    \centering
         \includegraphics[keepaspectratio=true, width=\linewidth]{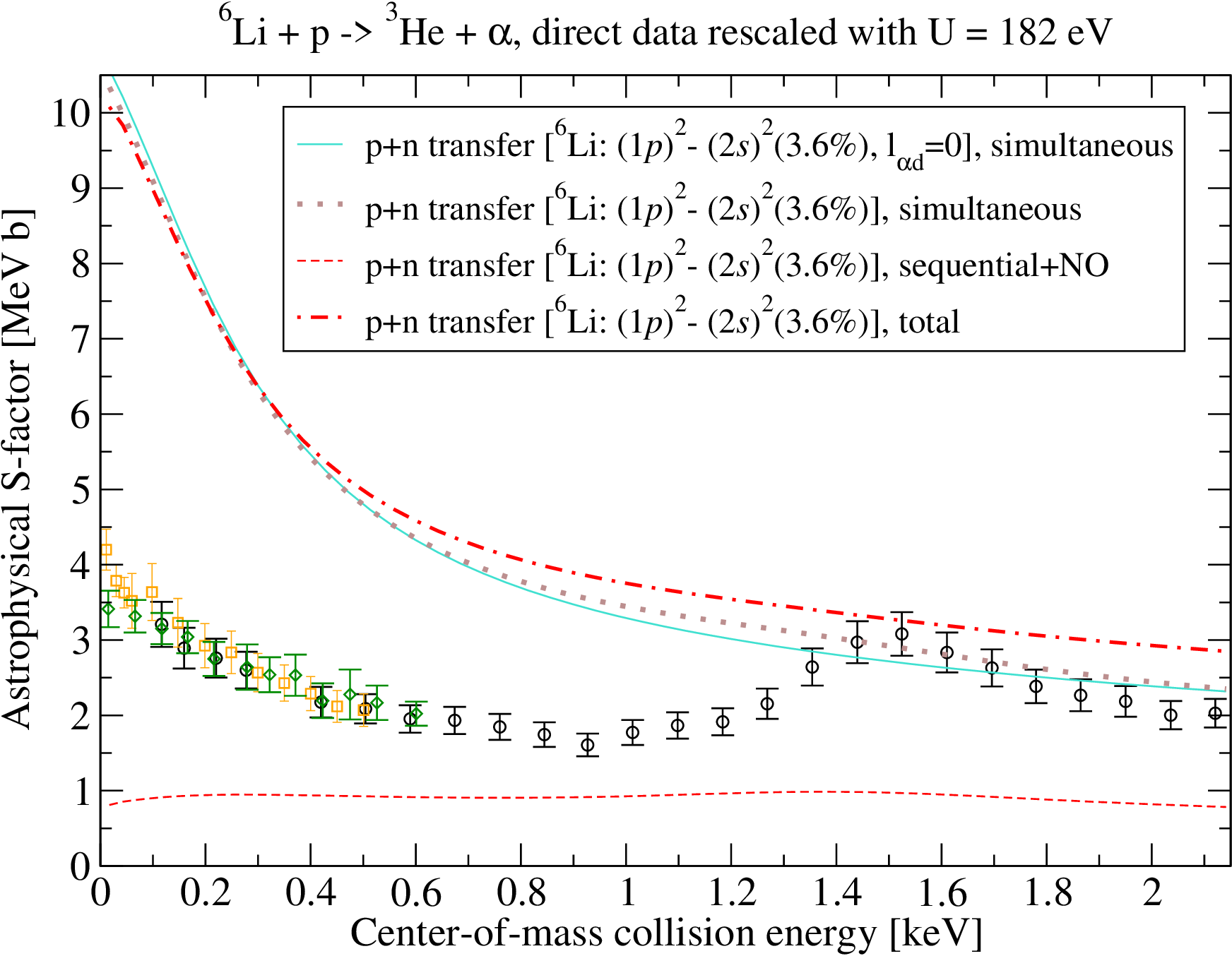}
    \caption{The red dot-dashed line represents the
    \nuclide[6]{Li}(\nuclide{p},\nuclide[3]{He})\nuclide{\alpha} astrophysical $S$ factor for the \npPairGroup{}-transfer DWBA calculation using the \nuclide[6]{Li} wave-function in the top-left panel of \cref{fig6LiTNTpdf}. The red dashed and brown dotted lines represent the $S$ factor associated with, respectively, only the first- and second-order contributions to the total transition amplitude, namely $\mathcal T^{(1)}$ and $\mathcal T^{(2)}$ in sec.~\ref{secMethodTwoParticleTransfer}.
    The turquoise solid line represents      the first-order-only $S$ factor, but excluding all contributions due to \nuclide[6]{Li} configurations where the \nuclide{\alpha}--\nuclide{d} relative-motion orbital angular momentum, $\vec\ctom_{\nuclide{\alpha}\nuclide{d}}$, is greater than 0 (see text for discussion).
    Points are the same experimental data in \cref{fig:dtransfer} (not shown in legend for brevity).}
    \label{fig:nptransfer}
\end{figure}
 The astrophysical $S$-factor  for \nuclide{p}+\nuclide{n} direct transfer obtained in  second-order DWBA  is represented in \cref{fig:nptransfer}.
It can be seen that the computed cross section overestimates the data, being higher
by about a factor 2 than the one obtained in the one-particle-transfer calculation (see \cref{fig:dtransfer}). The overestimation of the data seems to be a common problem with other microscopic calculations \cite{Arai2002,Vasilevsky2009}; in our two-nucleon-transfer calculations the discrepancy could be accentuated by the approximation employed for the evaluation of the transition potential in the simultaneous transfer scheme, see the last paragraph of sec.~\ref{secMethodTNTSimultaneous}, and by the use of a fictitious bound \nuclide[5]{Li} in the second-order calculation. 
However, as we will discuss in the following, the approach adopted here has the advantage of allowing to probe directly the link between clustering and the characteristics of 
the cross section; we will concentrate on this aspect hereafter.
      
 \Cref{fig:nptransfer} also includes  the $S$ factors  associated with the first-order simultaneous term only and the second-order term (including sequential and non-orthogonality, “NO”, terms) only.
   Decomposing the cross section in terms of the $^6$Li--\nuclide{p} initial orbital angular momentum, $\PTom_{\nuclide{Li}\nuclide{p}}$, we find that 
(not shown on the figure)
the total cross section at low energies is dominated by $\PTom_{\nuclide{Li}\nuclide{p}}=0$, as expected in general for non-resonant reactions in this regime \cite[sec.~4-5]{Clayton1983}.
  The $\PTom_{\nuclide{Li}\nuclide{p}}=0$ components of both  simultaneous and {second-order} processes are non-negligible, and interfere destructively.  The region of the resonance is dominated by $\PTom_{\nuclide{Li}\nuclide{p}}=1$ components, as expected in literature  (see e.g.~\ccite{Tumino2003}),
 mostly appearing in the sequential process.

Regarding the angular momentum decomposition of the structure wave-functions, there is now an additional degree of freedom with respect to the case in sec.~\ref{d results}. In “T" Jacobi coordinates, the wave-functions can be divided in components with definite relative orbital angular momentum between the two transferred nucleons, $\vec\nnom_{nn}$, and definite relative orbital angular momentum between the core particle (e.g.~\nuclide{\alpha}) and the center of mass of the transferred system, $\vec\ctom_{ct}$, see \cref{figCoordinate}.
The additional degree of freedom enlarges the set of allowed configurations. For instance, it is possible to form a component of \nuclide[6]{Li} with odd values of both $\vec\nnom_{nn}$ and $\vec\ctom_{\nuclide{\alpha}\nuclide{d}}$ (so that the state has the correct total parity).  The total orbital angular momentum,  which is conserved by the interactions in use, is the sum $\vec\nnom_{nn}+\vec\ctom_{ct}+\vec\PTom$ (with $\vec\PTom$  being the projectile-target orbital momentum).
\Cref{fig:nptransfer}  shows  a simultaneous-term calculation performed considering only 
$\vec\ctom_{\nuclide{\alpha}\nuclide{d}}=0$ for the \nuclide[6]{Li} configuration.
We observe that, at low energies, the latter is  close to the full simultaneous calculation, whereas the cross section above the barrier
is slightly  reduced, similarly to what was found in the deuteron-transfer case in sec.~\ref{secDeuteronTransferRoleDeformation}.
   
\begin{figure}[tbp]
    \centering
         \includegraphics[keepaspectratio=true, width=\linewidth]{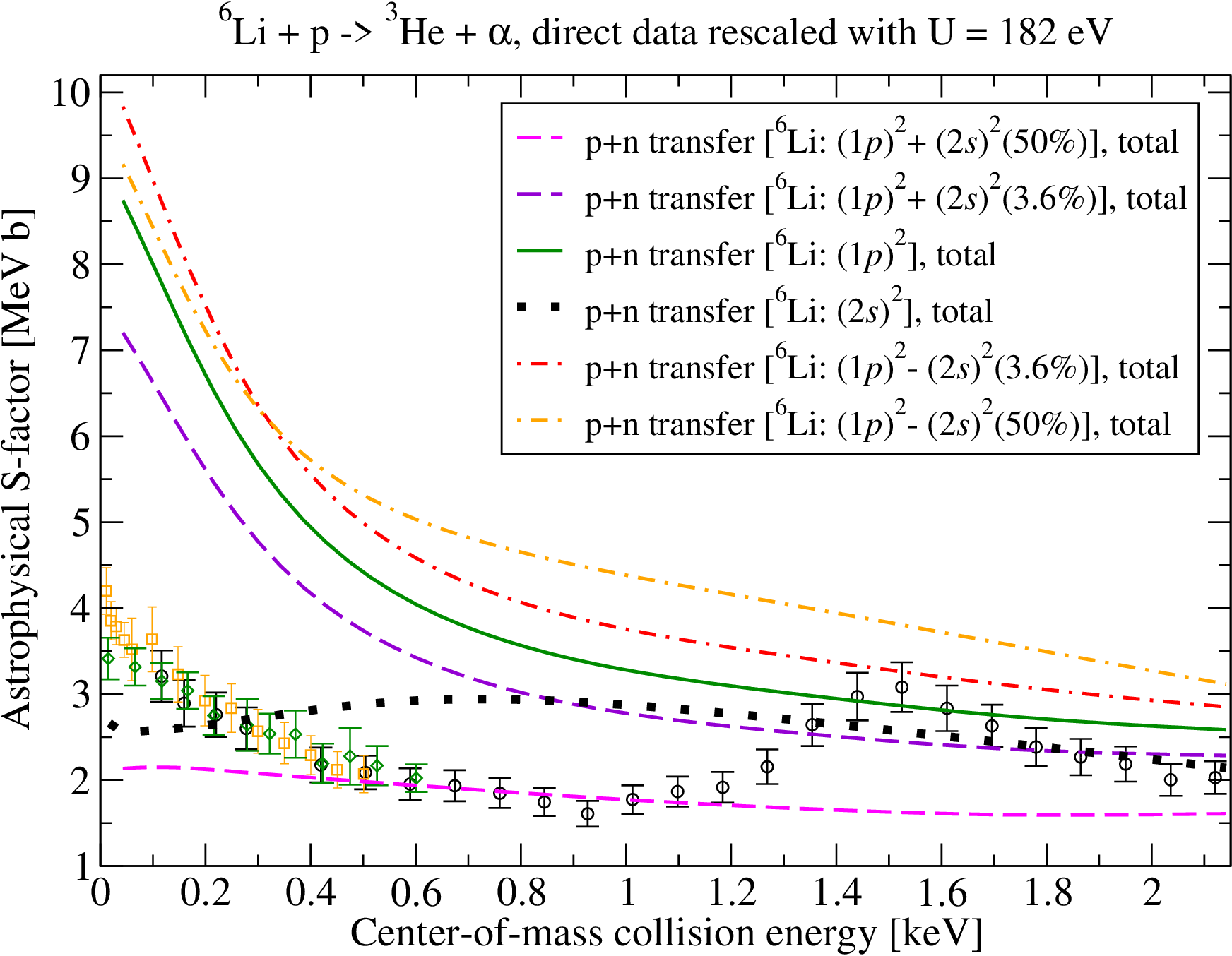}      \caption{     \npPairGroup{}-transfer astrophysical $S$ factors obtained for different choices of the $^6$Li wave-function,      as per the legend      (matching the labels in \cref{tabClusteredPeak,fig6LiTNTpdf}).      The green solid line corresponds to the middle-left panel of \cref{fig6LiTNTpdf}, red dot-dashed and violet dashed lines (“($2s$)$^2$(\qty{3.6}{\percent})" in the legend) correspond respectively to top- and bottom-left panels of \cref{fig6LiTNTpdf}, orange dot-dashed and magenta dashed lines (“($2s$)$^2$(\qty{50}{\percent})" in the legend) are associated with the top- and bottom-right panels of \cref{fig6LiTNTpdf}, while the black dotted line refers to the middle-right panel of \cref{fig6LiTNTpdf}. Experimental data are the same as in \cref{fig:dtransfer} (not shown in legend for brevity).}
                   \label{fig:nptransfer_clustering}
\end{figure}
      
\begin{figure}[tbp]
    \centering
         \includegraphics[keepaspectratio=true, width=\linewidth]{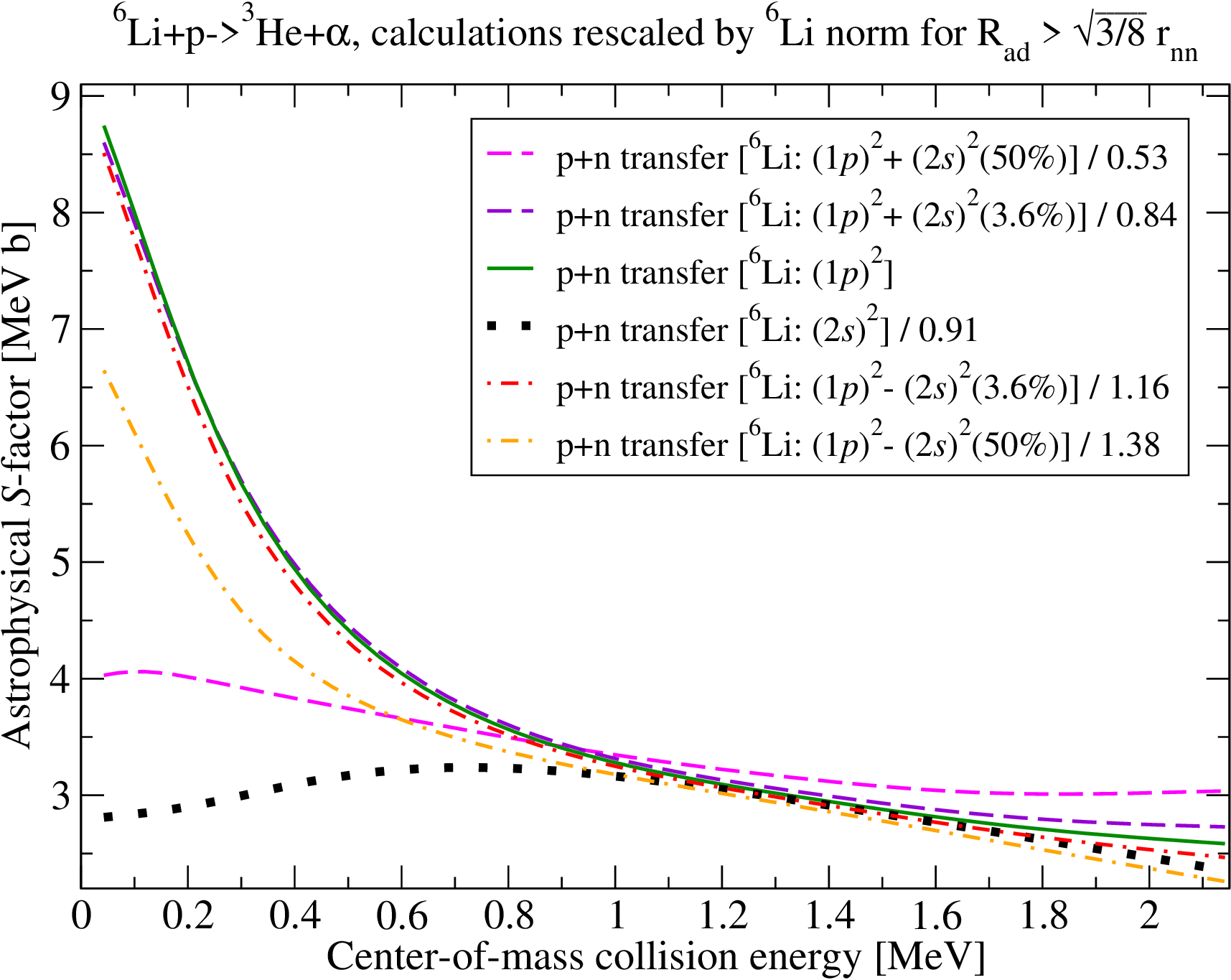}
    \caption{Same lines as in \cref{fig:nptransfer_clustering}, each divided by a constant factor proportional to the “clustered region" norm of the associated \nuclide[6]{Li} wave-function, as per \cref{tabClusteredPeak} (see sec.~\ref{secThreeparticlewavefunctions} for details), taking the pure $(1p)^2$ case as reference. 
         }
    \label{fig:nptransferRescalingNormClusteredPeak}
\end{figure}

\Cref{fig:nptransfer_clustering} illustrates results for the $S$ factor 
obtained with different options for the \nuclide[6]{Li} wave function (the same appearing in \cref{fig6LiTNTpdf}),  The cases leading to larger cross sections correspond to configurations  with a more pronounced clustered structure, that is, with larger norm in the ``clustered region'' (area above the red line in \cref{fig6LiTNTpdf}).
     Furthermore, we found that the absolute value of the computed cross-section, in the energy region around \qty{1}{\MeV}, scales with
 the amplitude of such norm,    as shown in \cref{fig:nptransferRescalingNormClusteredPeak}.    
 This scaling appears to be a clear sign of the role of the clustering strength within the transferred system in the direct reaction process.
However, the figure evidences that the low-energy trend is affected by more specific features of the structure wave function; a cross-section enhancement in this region appears to be
favored for the configurations exhibiting a greater maximal probability 
  in the “clustered region", which is reported in \cref{tabClusteredPeak} for each tested \nuclide[6]{Li} wave function. As can be seen by comparison with \cref{fig:nptransferRescalingNormClusteredPeak}, such “clustered peak height" appears to be correlated with the relative excursion of the astrophysical factor between \qty{1.2}{\MeV} and the lowest explored energies.
In particular, configurations with a larger $(1p)^2$ component (left panels of \cref{fig6LiTNTpdf}), leading to a well pronounced “clustered" peak, display a steeper low-energy trend.
    We mention that other scaling trends could be found,  considering for instance
the root-mean-square radius of an effective \nuclide{\alpha}--\nuclide{d} probability density function, obtained either integrating the \nuclide[6]{Li} probability density function, or from
 the projection of  the complete \nuclide[6]{Li} wave-function on the free-deuteron ground state (representing a strongly-clustered configuration), computed similarly in \cite[sec.~5]{Bang1979}.  We have checked that the norm of this latter projection also correlates with the cross-section absolute values.
     The observation of scaling trends on several benchmarks   is mainly an indication of the correlation between the aforementioned properties of the wave-function.

\begin{figure}[tbp]
    \centering
    \includegraphics[keepaspectratio=true, width=\linewidth]{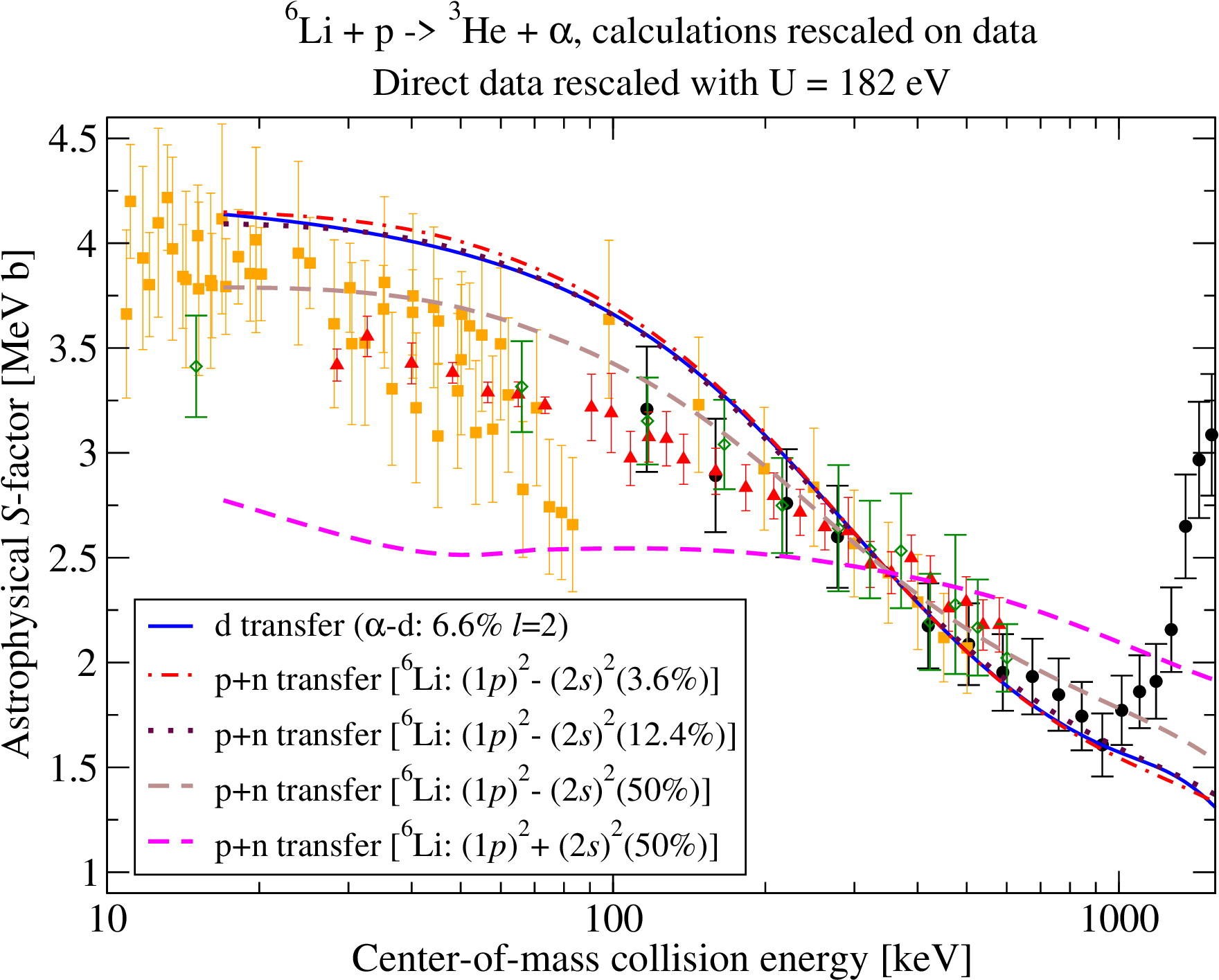}
    \caption{Points are the same rescaled data      in the bottom panel of \cref{fig1} (not shown in legend for brevity).      Lines are \nuclide{d}- and \npPairGroup{}-transfer calculations obtained for different choices of $^6$Li wave-function, as per the legend (matching the labels in      \cref{fig:dtransfer,fig:nptransfer_clustering,tabClusteredPeak}),
              but each was multiplied by a distinct constant to match experimental data at \qty{350}{\keV}.}
     \label{fig10}
\end{figure}

\Cref{fig10} compares selected results related to the two-nucleon and inert-deuteron transfer. In particular, the two-nucleon-transfer astrophysical factor for three of the \nuclide[6]{Li} wave-functions in \cref{tabClusteredPeak} is compared to   the complete  deuteron-transfer  calculation in \cref{fig:dtransfer} adopting the strongest $\ctom_{\nuclide{\alpha}\nuclide{d}} = 2$ component for the \nuclide[6]{Li} state (blue line).    All curves are rescaled by a distinct constant adjusted on experimental data. 
  We observe that the deuteron-transfer calculation has a low-energy trend similar to those of the aforementioned two-nucleon transfer calculations considering a large $(1p)^2$ contribution (featuring the most pronounced clustered peak). This appears to be consistent with the idea that, in the \nuclide{d}-transfer case, the deuteron internal state is decoupled and frozen to the  the free-deuteron one.
  This observation nicely supports the occurrence of clustering in the ground state configuration as a possible candidate to explain the behavior of the astrophysical factor at very low energies and the so-called electron screening puzzle.  
On the other hand, the calculations corresponding to the \qty{50}{\percent} mixing of $(1p)^2$ and (2s)$^2$ configurations clearly show a flatter trend at low energy, especially in the least-clustered case (“$1p + 2s(50\%)$" in the figures legend), as already observed in \cref{fig:nptransferRescalingNormClusteredPeak}.

\section{Summary and perspectives}\label{secConclusions}
We have presented an analysis of the \nuclide[6]{Li}(\nuclide{p},\nuclide[3]{He})\nuclide{\alpha}
transfer reaction, based on first- and second-order DWBA calculations. Our main motivation
is linked to the analysis of the cross section at energies of astrophysical 
interest, also in connection with the quite debated anomalous enhancement 
observed in several sets of experimental data. 
It has been recently proposed that this observation could be ascribed to
clustering effects \cite{Spitaleri2016}, possibly inducing also deformation effects  in the 
ground state configuration of the involved nuclei.   
Along this trail, we have performed DWBA calculations adopting two {distinct models}:
we consider the transfer of a single inert particle, namely a deuteron, or we
adopt the more complex picture of a \npPair{} transfer, allowing for different 
possible configurations of the two nucleons with respect to the $\alpha$-particle core.

The deuteron-transfer calculation leads to a reasonable reproduction of the measured transfer cross section at very low energies. It is interesting to note that
our present DWBA results are quite close to those obtained from past resonating-group-method calculations (see \cref{fig:dtransfer}).
One can argue that the use of sufficiently rich projectile-target interactions, capturing essential properties of interest of the elastic scattering process, contributed ensuring that the computed transfer cross-section was free from spurious resonances, thus increasing the reliability of the results. 
Moreover, within our approach, it was also possible to investigate the role of deformed components in the reactants wave-functions (see sec.~\ref{secDeuteronTransferRoleDeformation}).

The \npPair{}-transfer picture is more flexible. It allows to gauge the degree of clustering and to probe
its impact on the features of the transfer cross section. 
In all cases, we observe that a 
clustered configuration of the \npSystem{} system, characterized  by a short relative distance between the two nucleons, 
favors the transfer mechanism. We also find that the cross-section {overall magnitude}  
   approximately scales with the integral of the \nuclide[6]{Li} probability 
density function
over the region above the red line in each panel in \cref{fig6LiTNTpdf}, populated by more strongly clustered configurations.
     The \nuclide[6]{Li} wave-function structures predicted by microscopic calculations \cite{Bang1979,Casal2021Private}, which are characterized by a relevant weight of $(1p)^2$ configurations (as generally expected for the ground state), are instead particularly effective in enhancing the astrophysical factor at very low energies.

Moreover, our calculations show that the enhancement of the low-energy cross
section is not   ascribable to static deformation effects of the ground state
 but comes from the presence of clustered components in the wave function.   We note that 
 the wave functions
adopted or emerging
from our calculations include only relatively small contributions from non-spherical waves in the core-deuteron (or \npSystem{} system) relative motion.
     
As a drawback, the approach adopted here leads to a global overestimation of the experimental data, which seems to be a common
problem with microscopic 
approaches \cite{Arai2002,Vasilevsky2009}. The issue is particularly relevant in the two-nucleon-transfer calculations, where it might originate from the approximations employed in applying state-of-the-art two-nucleon-transfer numerical methods to the light systems considered
here. In perspective, in order to address this problem it could be beneficial to consider more microscopic (three-body) approaches for the description of the \nuclide[6]{Li} and \nuclide[3]{He} structure wave-functions \cite{Bang1979,Tanihata2008,Casal2013}.   Finally, in the present work we have considered possible clustering and deformation effects 
only in the initial and final (ground-state) configurations of the involved systems.  
The discussion of dynamical deformation or reorientation effects, namely polarization effects, 
and of their possible impact on the transfer cross section, will be the 
subject of future work.

To overcome the drawbacks discussed above, this investigation
could possibly be approached from a more ab-initio perspective. Resonating-group-method calculations might be improved within the no-core shell-model with continuum framework \cite{Navratil2016}. A microscopic four-body treatment (e.g., Faddeev-Yakubovsky \cite{Lazauskas2020} or Alt-Grassberger-Sandhas \cite{Deltuva2014,Deltuva2017,Watanabe2021}) of the \nuclide{\alpha}+\nuclide{p}+\nuclide{n}+\nuclide{p} system (treating the \nuclide{\alpha}-particle as inert) may also be feasible.
A comparison with a four-body Faddeev AGS calculation, where in principle the same interactions can be used, will be of great help to understand whether the disagreement in the absolute value with the experimental data comes from limitations regarding the DWBA approximation or from the interactions used. A similar comparison has found a satisfactory agreement for $(\nuclide{p},\nuclide{p}N)$ transfer reactions \cite{GomezRamos2020}, but no benchmark has been done at energies of astrophysical interest to our knowledge.

\appendix
 \section{Adopted potentials}\label{secPotentialsDefinition}
 \begin{table*}[tbp]  		\caption{\label{tabPotPar}              List of all potentials employed in this work.
			Each column refers to a different potential, identified in the column header by the pair of particles it refers to.  			Each line refers to a parameter in \cref{eqGeneralPotentialParametrization} (with same notation, {see text for details}).
			 			 			 			 			         ``{\nuclide{p} -- \nuclide{d} ($L=0$)}'' or ``$L=2$'' refers to the two potentials employed in sec.~\ref{secResults} to construct the \nuclide[3]{He} state.}
		\centering
	 		\sisetup{table-format=2.4}
		\begin{tabular}{lSSSSS[table-format=3.4]S[table-format=+3.4]S[table-format=+4.4]SSS}  & {\nuclide[6]{Li} -- \nuclide{p}} & {\nuclide{\alpha} -- \nuclide[3]{He}} & {\nuclide[5]{Li} -- \nuclide{d}} & {\nuclide[5]{Li} -- \nuclide{p}} & {\nuclide{p} -- \nuclide{n}} & {\nuclide{p} -- \nuclide{d} ($L=0$)}  & {\nuclide{p} -- \nuclide{d} ($L=2$)} & {\nuclide{\alpha} -- \nuclide{d}} & {\nuclide{\alpha} -- \nuclide{p}} & {\nuclide[5]{Li} -- \nuclide{n}} \\ \toprule
\vphantom{$\left(\nuclide[6]{Li}\right)$}$R_C$ [\si{\femto\metre}]	&  2.326  &  0.1000 &  2.223  &  2.233 & {--} & 2.000 & 2.000 & 1.900 & 2.900 & {--} \\  			\midrule
\vphantom{$\left(\nuclide[6]{Li}\right)$}$V_v$ [\si{\MeV}]			& 48.20   & 66.08   & 90.04   & 50.97   & 165.4  & 179.9 & 8155 & 80.09 & 43.00 & 43.00 \\  \vphantom{$\left(\nuclide[6]{Li}\right)$}$R_v$ [\si{\femto\metre}]	&  1.908  &  2.649  &  2.001  &  1.912  &  0.4000  & 0.5400 & -2.190 & 1.900 & 2.000 & 2.125 \\  \vphantom{$\left(\nuclide[6]{Li}\right)$}$a_v$ [\si{\femto\metre}]	&  0.6700 &  0.7175 &  0.7090  &  0.6900  &  0.6000  & 0.6800 & 0.9100 & 0.6500 & 0.7000 & 0.7000 \\
			\midrule
\vphantom{$\left(\nuclide[6]{Li}\right)$}$V_g$ [\unit{\MeV}]        & 0  & 0  & 0  & 0  & 0 & -203.3 & -8400 & 0 & 0 & 0 \\
\vphantom{$\left(\nuclide[6]{Li}\right)$}$a_g$ [\si{\femto\metre}]	& {--} & {--} & {--} & {--} & {--} & 0.6400 & 0.3500 & {--} & {--} & {--} \\
            \midrule
\vphantom{$\left(\nuclide[6]{Li}\right)$}$V_w$ [\si{\MeV}]			&  0	  &  0		& 0		  &  0.6908  & 0  & 0  & 0 & 0 & 0 & 0 \\
\vphantom{$\left(\nuclide[6]{Li}\right)$}$R_w$ [\si{\femto\metre}]	& {--}	  & {--}	& {--}	  &  1.854 & {--}   & {--} & {--} & {--} & {--} & {--} \\
\vphantom{$\left(\nuclide[6]{Li}\right)$}$a_w$ [\si{\femto\metre}]	& {--}	  & {--}	& {--}	  &  0.6900 & {--}  & {--} & {--} & {--} & {--} & {--} \\
			\midrule
\vphantom{$\left(\nuclide[6]{Li}\right)$}$V_x$ [\si{\MeV}]			&  0.2246 &  1.089  & 12.20	  &  4.723  & 0  & 0  & 0 & 0 & 0 & 0 \\  \vphantom{$\left(\nuclide[6]{Li}\right)$}$R_x$ [\si{\femto\metre}]	&  3.634  &  2.102  &  2.266  &  1.854  & {--} & {--} & {--} & {--} & {--} & {--} \\  \vphantom{$\left(\nuclide[6]{Li}\right)$}$a_x$ [\si{\femto\metre}]	&  2.715  &  0.7386 &  0.6497 &  0.6900  & {--} & {--} & {--} & {--} & {--} & {--} \\
			\midrule
\vphantom{$\left(\nuclide[6]{Li}\right)$}$V_o$ [\si{\MeV}]			&  1.000  &  4.162  &  7.330  &  0  & 0  & 0  & 1.470 & 0 & 10.00 & 10.00 \\
\vphantom{$\left(\nuclide[6]{Li}\right)$}$R_o$ [\si{\femto\metre}]	&  1.817  &  2.644  &  1.830  &  {--} & {--} & {--} & 2.070 & {--} & 1.500 & 1.594 \\
\vphantom{$\left(\nuclide[6]{Li}\right)$}$a_o$ [\si{\femto\metre}]	&  0.700  &  0.2078 &  0.6600 &  {--} & {--} & {--} & 0.0600 & {--} & 0.3500 & 0.3500 \\
			\midrule
\vphantom{$\left(\nuclide[6]{Li}\right)$}$V_s$ [\si{\MeV}]			& 18.00   & {--} & 0 & 0 & 0  & 0  & 0 & {--} & {--} & 0 \\
\vphantom{$\left(\nuclide[6]{Li}\right)$}$R_s$ [\si{\femto\metre}]	&  1.853  & {--} & {--} & {--} & {--} & {--} & {--} & {--} & {--} & {--} \\
\vphantom{$\left(\nuclide[6]{Li}\right)$}$a_s$ [\si{\femto\metre}]	&  0.200  & {--} & {--} & {--} & {--} & {--} & {--} & {--} & {--} & {--}
		\end{tabular}
	 \end{table*}
     
                                         The two-body potentials employed in this work were parameterized as follows:
   	                \begin{equation}
	\begin{aligned}\label{eqGeneralPotentialParametrization}
	\mathcal V&_{1,2}(r) = \\
    & V_C(r, R_C) - V_v \, f(r,R_v,a_v) - V_g \, \textrm{e}^{-(r/a_g)^2}      + \\
    & - i V_w \, f(r,R_w,a_w)
	+ i 4 V_x a_x \, \frac{\d}{\d r} f(r,R_x,a_x) + \\
    & + 2 \, \vec l \cdot \vec s_1 V_o \frac{\SI{2}{\femto\metre\squared}}{r} \, \frac{\d}{\d r} f(r,R_o,a_o) + \\
	& + 2 \, \vec s_1 \cdot \vec s_2 V_s \, f(r,R_s,a_s) ,
	\end{aligned}\end{equation}
	 	where:
	 	\begin{equation}\label{eqCoulombWoodsSaxonDefinition}\begin{gathered}
	V_C(r, R_C) = k_e   Z_1 Z_2 \left\lbrace\begin{aligned}
 		&\frac{ 3 - r^2 / R_C^2 }{2 R_C} & r < R_C \\
		&\frac{1}{r} & r \geq R_C
		\end{aligned}\right. ,
	\\ 	f(r,R,a) = \frac{1}{1 + \exp\left(\frac{r-R}{a}\right)},
	\end{gathered}\end{equation}
	     $k_e \approx \qty{1.44}{\MeV \femto\metre}$,
    $Z_1$ and $Z_2$ are the particles      charge numbers,
     and “$2 \, \vec a \cdot \vec b$" is a shorthand for $c (c+1) - a (a+1) - b (b+1)$, where $c$ is the modulus quantum number of the coupling of the angular momenta associated to $a$ and $b$.
    $r$ represents the distance between the interacting particles, while $s_1$, $s_2$ and $l$ are the modulus quantum numbers for, respectively, the particles intrinsic spin and their relative orbital angular momentum.
          	          In all potentials taken from literature,  	the spin-orbit term couples only the lightest particle spin.      Only in the fitted \nuclide[6]{Li}--\nuclide{p} potential, the spin-orbit term involves the total intrinsic spin (coupling of $\vec s_1$ and $\vec s_2$), whose modulus quantum number is denoted by $s$: in this case, the potential depth has thus the form $2 \, \vec l \cdot \vec s \, V_o$,      differently      from what is stated in \cref{eqGeneralPotentialParametrization}.       	  \Cref{tabPotPar} lists the parameters value adopted for each potential.
 Whenever a potential was employed to construct a bound wave-function, $V_v$ was rescaled to obtain the desired binding energy for the system. Whenever a potential was employed as core-core interaction, all non-central terms were discarded.
  
\begin{acknowledgments}
We warmly thank
C.~Spitaleri, A.~Moro, L.~Fortunato, S.~Typel, and E.~Vigezzi for inspiring discussions.
We also acknowledge the ASFIN collaboration for useful discussions on their recent data.
 This project has received funding from the European Research Council (ERC) under the European Union's Horizon 2020 research and innovation program (grant agreement No.~714625).
This work is based on research supported in part by grants PID2020-114687GB-I00 and FIS2017-88410-P funded by MCIN/AEI/10.13039/501100011033,
grant Group FQM-160 and the project PAIDI 2020 with Ref. P20\_01247, funded by the Consejer\'{\i}a de Econom\'{\i}a, Conocimiento, Empresas y Universidad, Junta de Andaluc\'{\i}a (Spain), and by ERDF A way of making Europe.
\end{acknowledgments}
\bibliography{bib}
\end{document}